# Extracting effective scaling exponents in finite-size hyperuniform systems

Yuan Liu [1*], Xurui Li [1*], Jianxiang Tian [12†], Xunwang Yan[1†] and Ge Zhang[3†]

[1]*Department of Physics, Qufu Normal University, Qufu, 273165, PR China*
[2]*Department of Physics, Dalian University of Technology, Dalian, 116024, PR China*
[3]*Department of Physics, City University of Hong Kong, Kowloon, 518057, Hong Kong China*
**These authors contributed equally to this work*
[†]*Corresponding authors*

jxtian@qfnu.edu.cn

xwyan@qfnu.edu.cn

gzhang37@cityu.edu.hk

**Abstract:** Hyperuniform systems strongly suppress long-wavelength density fluctuations, which is quantitatively characterized by the small-wavenumber scaling $S(k) \sim k^{\alpha}$. In finite samples, however, accurately estimating the hyperuniformity exponent $\alpha$ can be challenging. The inferred value depends strongly on the range of length scales accessible in the measurement, finite-size effects, and the specific characterization method employed, whether based on Fourier-space structure factors, real-space density fluctuations, or dynamical probes such as diffusion spreadability. In particular, the structure-factor method provides the most direct estimate of $\alpha$, but is sensitive to empirical low-$k$ fitting cutoffs. The number-variance method offers a real-space Class-like diagnosis, but contributes a numerical exponent only when the finite-size data retain Class III-like scaling information. The spreadability method provides a smoother dynamic estimate and reduces configuration-level fluctuations, but requires a physically admissible long-time fitting window. Here, we develop a practical method-aware protocol for robust estimation of the effective scaling exponent $\alpha$ in finite-size hyperuniform point configurations, combining three complementary methods with distinct roles. Our protocol summarizes the method-specific estimates through a joint empirical estimator and reports the internal dispersion among the participating methods to determine the optimal estimate. We benchmark the protocol on two-dimensional targeted hyperuniform point configurations generated under prescribed low-$k$ structure-factor constraints for $\alpha_{\text{theory}}$=0.3, 0.5, 0.7, 1.0, 1.5, 2.0, 2.5, 3.0 and 4.0. We find that the regularized $S(k)$ fit reduces sensitivity to empirical cutoffs, the number-variance method separates finite-size Class-like regimes, and the plateau-window spreadability analysis suppresses configuration-level variability. Across the benchmark set, the joint estimates remain close to the prescribed exponents and provide reproducible finite-size effective summaries. The framework offers a method-aware procedure for comparing finite hyperuniform samples and diagnosing where individual methods retain or lose exponent information.

## I. INTRODUCTION

Hyperuniform systems are many-body systems in which density fluctuations are anomalously suppressed in the long-wavelength limit. They include perfect crystals, perfect quasicrystals, and certain disordered states, and provide a unified framework for characterizing large-scale order in point configurations and heterogeneous media [1-4]. Research on hyperuniformity has developed into a broad interdisciplinary field, with applications and realizations in photonic and phononic band-gap materials [5-9], antenna and laser design [10], stealthy hyperuniform materials [11-12], transport phenomena and superconducting systems [13], diffusion in two-phase media [14-17], condensed-matter systems [18], mathematical point sets [19-22], network systems and network materials [23-27], biological tissues [28-37], spin systems, active matter, and nonequilibrium systems [38-40], anisotropic random fields [41], and jammed packings and contact-number hyperuniformity [42-44]. Hyperuniform disordered systems have also been extensively studied in jammed and glassy packings, where suppressed long-wavelength density fluctuations emerge without crystalline order [45]. These studies show that hyperuniformity is not only a structural classification, but also a useful concept for linking microscopic organization to large-scale physical response. A quantitative question that naturally follows is how to characterize not only whether a system is hyperuniform, but also the manner in which its long-wavelength fluctuations vanish.

For point configurations, the standard characterization of hyperuniformity is based on the static structure-factor $S(k)$, which measures density fluctuations in reciprocal space [1, 4]. A statistically homogeneous point configuration is hyperuniform if

$$\lim_{k\to 0} S(k) = 0 \tag{1}$$

This condition means that density fluctuations are suppressed in the infinite-wavelength limit. In many theoretical models, numerical configurations, and experimental systems, the small-wavenumber behavior can be further described by a power law,

$$S(k) \sim k^{\alpha} \qquad (k \to 0) \tag{2}$$

where $\alpha$ is the scaling exponent. The exponent specifies how rapidly $S(k)$ approaches zero at small $k$. It therefore provides more information than a binary distinction between hyperuniform and non-hyperuniform behavior.

The value of $\alpha$ is important because different exponents correspond to different large-scale fluctuation laws. In this sense, the central question is not only whether a system is hyperuniform, but also how its long-wavelength fluctuations vanish [3-4, 46]. The exponent connects reciprocal-space scaling, real-space number fluctuations, and, for two-phase media, dynamic diffusion response [14-17]. Reliable extraction of $\alpha$ from structural data is therefore a basic task in the quantitative analysis of hyperuniform systems.

Existing approaches to $\alpha$ can be organized into three complementary methods. The first method is the direct structure-factor method. Since $S(k)$ directly measures reciprocal-space density fluctuations, one may in principle estimate $\alpha$ by fitting the low-$k$ behavior in Eq. (2). This method is the most direct one, but it is also sensitive to the number of available low-$k$ points and to the selected fitting window.

The second method is based on real-space number fluctuations. For an observation window of radius $R$, the number variance is defined as [1, 3]

$$\sigma_N^2(R) = \langle N(R)^2 \rangle - \langle N(R) \rangle^2 \tag{3}$$

where $N(R)$ is the number of points inside the window and the angular brackets denote an average over window positions. This quantity measures the fluctuation of local particle number as the window is moved through the system. In two dimensions, the asymptotic growth of $\sigma_N^2(R)$ is directly related to the small-$k$ exponent. For $0 < \alpha < 1$, one has $\sigma_N^2(R) \sim R^{2-\alpha}$. For $\alpha = 1$, the leading behavior becomes $R \ln R$. For $\alpha > 1$, the dominant growth is proportional to $R$ [1, 3-4, 46]. Thus, number variance provides a real-space method for Class-like diagnosis and, in the Class III-like regime, can also provide approximate exponent information.

The third method is based on spreadability in two-phase media. For a two-phase medium, $S(t)$ measures the fraction of diffusing mass that has entered the target phase at time $t$, while the spreadability excess $S(\infty) - S(t)$ measures the remaining deviation from the long-time limit. Theoretical work has shown that spreadability is connected to the two-point statistics of the medium. If the relevant spectral density satisfies

$$\tilde{\chi}_V(k) \sim |k\ell|^\alpha \qquad (k \to 0) \tag{4}$$

then the long-time decay of the spreadability excess follows [15-16]

$$S(\infty) - S(t) \sim \left(\frac{Dt}{\ell^2}\right)^{-(d+\alpha)/2} \tag{5}$$

where $D$ is the diffusion coefficient, $\ell$ is a characteristic length scale, and $d$ is the spatial dimension. For point configurations, this method can be applied by decorating the points with identical disks and defining the target phase as the union of the disk domains [17, 47]. In the low-area-fraction decoration used here, the resulting two-phase medium provides a dynamic probe of the large-scale spectral behavior encoded by the underlying point pattern. The spreadability excess therefore supplies a complementary finite-size method to the same structural information analyzed through $S(k)$.

The three complementary methods described above characterize the same large-scale structural problem from different perspectives: reciprocal space, real space, and dynamic diffusion response. In the thermodynamic limit, their asymptotic forms are theoretically connected. In finite simulations and finite experimental fields of view, however, the required limiting regimes are not directly accessible. The smallest available wavenumber is set by the system size. The largest reliable observation window for number variance is limited by the finite domain. The long-time spreadability regime is also restricted because the diffusion length eventually becomes comparable to the simulation box [48-50]. Finite-window sampling, random spatial truncation, and reciprocal-space binning can therefore influence both hyperuniformity detection and the measured exponent in finite data [48].

Several finite-size diagnostics have been proposed to address the difficulty of applying an ideal thermodynamic-limit criterion to finite samples. One commonly used measure is the hyperuniformity index [1, 51-53]

$$H = \frac{S(0)}{S(k_{\mathrm{peak}})} \tag{6}$$

where $S(0)$ is obtained by extrapolating low-$k$ data and $k_{\mathrm{peak}}$ denotes the position of a principal peak in $S(k)$. A small value of $H$ is often used as an empirical indicator of effective hyperuniformity. From a real-space viewpoint, finite-size or nearly hyperuniform behavior can also be interpreted through the large-$R$ form [1, 53]

$$\sigma_N^2(R) \approx AR^d + BR^{d-1}, \qquad A = \lim_{k\to 0} S(k) \tag{7}$$

When $A$ is small but not strictly zero, the surface-area term $BR^{d-1}$ can dominate over an intermediate range of $R$. A finite sample may then show apparent hyperuniform-like behavior over a limited observation window, even though the strict infinite-system condition is not directly verified.

These considerations clarify the finite-size nature of the present problem. Existing diagnostics have shown that finite samples should often be discussed in terms of effective hyperuniformity rather than strict thermodynamic-limit hyperuniformity. What remains less developed is a unified procedure for extracting an effective scaling exponent across the structure-factor, number variance, and spreadability methods. In finite-size hyperuniform systems, exponent extraction is therefore not only a regression problem. It is also a window-selection problem and a method-consistency problem.

In this work, we focus on this methodological issue: how to extract a reproducible finite-size effective scaling exponent from hyperuniform data under a unified analysis protocol. Here, the finite-size effective scaling exponent is defined as a reproducible descriptor obtained for a specified system size, analysis protocol, and set of observable windows. It is not intended to approximate the thermodynamic-limit exponent by extrapolation. Its primary role is to provide a common basis for comparing finite samples analyzed under the same protocol and for diagnosing where different methods retain or lose exponent information.

To address this problem, we develop a unified finite-size protocol that assigns a specific role to each method. For the structure-factor method, we introduce regularized low-$k$ window screening. Instead of fitting a fixed number of the lowest-$k$ points or imposing a single empirical cutoff, the protocol scans contiguous low-$k$ windows and ranks them by a normalized score that combines log-log fitting quality with boundary perturbation stability. This converts the low-$k$ fitting problem into a reproducible finite-window selection problem.

For the number variance method, we use the large-$R$ behavior mainly for Class-like diagnosis. This treatment follows from the asymptotic number variance classification associated with $S(k) \sim k^{\alpha}$ [1, 3-4, 46]. In the Class III-like regime, the leading behavior $\sigma_N^2(R) \sim R^{2-\alpha}$ still carries exponent information and can provide an approximate finite-size reference for $\alpha$. In the Class II-like and Class I-like regimes, however, the leading forms $R\ln R$ and $R$ make number variance much less sensitive to different values of $\alpha > 1$. In these cases, number variance is used as a diagnostic method rather than as a primary exponent estimator.

For the spreadability method, we introduce a plateau-window protocol for spreadability excess $S(\infty) - S(t)$. The use of spreadability as a dynamic structural probe is based on its connection to the spectral density of two-phase media [15-17]. Although the spreadability curve is often smoother than the raw low-$k$ structure-factor, exponent extraction still depends on selecting an appropriate long-time window. Our protocol restricts the search by a diffusion-length constraint, evaluates candidate windows using both log-log fitting quality and local effective-exponent stability, and applies a shared fitting window to all configurations in the same ensemble.

The method-specific outputs are then combined through a joint empirical estimator. The $S(k)$ and spreadability methods provide the main numerical inputs. The number variance method contributes to the numerical average only when its Class-like diagnosis indicates that exponent information is available. We also report the internal consistency $u_{\text{joint}}$, which measures the dispersion among the participating method estimates. This quantity is not a statistical confidence interval or an asymptotic error bar. It only reports the internal agreement among the finite-size methods under the same analysis protocol.

As a controlled benchmark, we use two-dimensional target hyperuniform point configurations generated under prescribed $S(k)$ constraints [54]. The prescribed exponent $\alpha_{\text{theory}}$ defines the target low-$k$ scaling used during configuration generation. This benchmark provides a controlled setting for testing finite-size window selection, method complementarity, and joint-estimator stability across representative hyperuniformity classes.

The remainder of this paper is organized as follows. Section II describes the construction of the benchmark finite-size hyperuniform systems and clarifies the meaning of the target exponent $\alpha_{\text{theory}}$. Section III analyzes the main finite-size difficulties faced by the structure-factor, number variance, and spreadability methods. Section IV presents the proposed extraction framework, including regularized low-$k$ window screening, Class-like diagnosis based on number variance, the plateau-window protocol for spreadability excess $S(\infty) - S(t)$, and the joint empirical estimator. Section V applies the framework to the benchmark systems and compares the method-specific and joint estimates. Section VI summarizes the main conclusions and discusses possible extensions to experimental and more complex finite-size data.

## II. BENCHMARK FINITE-SIZE HYPERUNIFORM SYSTEMS

To test the stability and reproducibility of effective-exponent extraction under finite-size constraints, we use a controlled set of two-dimensional target hyperuniform point configurations. The purpose of this benchmark is methodological. We do not aim to determine whether a particular interaction model is strictly hyperuniform in the thermodynamic limit. Instead, we ask how an effective scaling exponent can be extracted from finite samples, finite fitting windows, and a fixed analysis protocol. A suitable benchmark should therefore satisfy three requirements. The target low-$k$ exponent should be prescribed in advance. The system size and the number of configurations should be controlled. The tested exponents should span the main Class-like regimes so that the capabilities of the structure-factor, number variance, and spreadability methods can be compared under the same finite-size setting.

We generate the benchmark configurations using the target structure-factor construction described in Ref. [54]. In this approach, an ensemble of configurations is optimized so that the ensemble-averaged structure-factor follows a prescribed target form over a constrained low-wavenumber range. Specifically, the target structure-factor is chosen as

$$S_0(k) = \left(\frac{k}{k_{\max}}\right)^{\alpha_{\text{theory}}} \qquad |\boldsymbol{k}| < k_{\max} \tag{8}$$

where $k_{\max}$ defines the upper bound of the constrained spectral range. The exponent $\alpha_{\text{theory}}$ is the prescribed target exponent used during configuration generation. It is not an exponent obtained by fitting the finite samples, nor is it treated as a thermodynamic-limit exponent recovered from the data. During generation, the deviation between the ensemble-averaged $S(k)$ and the target $S_0(k)$ is minimized over the constrained range $|\boldsymbol{k}| < k_{\max}$. The resulting configurations therefore provide controlled finite-size samples whose ensemble-averaged low-$k$ behavior is designed to follow the specified target scaling within the imposed spectral window. In the present benchmark, $k_{\max}$ is fixed at 5 for all target exponents.

The main benchmark set contains nine target exponents: $\alpha_{\text{theory}}$ =0.3, 0.5, 0.7, 1.0, 1.5, 2.0, 2.5, 3.0, and 4.0. These values are not intended to exhaust all possible exponents. They are chosen to cover representative Class-like regimes in two dimensions. For $0 < \alpha_{\text{theory}} < 1$, the number variance is expected to retain leading exponent information through the large-$R$ behavior $\sigma_N^2(R) \sim R^{2-\alpha}$. The case $\alpha_{\text{theory}} = 1$ corresponds to the logarithmic crossover associated with $R\ln R$. For $\alpha_{\text{theory}} > 1$, the leading number variance growth in two dimensions is dominated by the $R$ term, so the number variance method is expected to provide mainly Class-like diagnosis rather than a precise numerical inversion of $\alpha$. This range of target exponents therefore allows us to examine low-$k$ window selection in the $S(k)$ method, long-time window selection in the spreadability method, and the capability boundary of the number variance method within one controlled framework [1, 3-4, 46].

Unless otherwise stated, the default benchmark uses $N_p = 200$, $N_c = 100$ and $\rho = 1.0$ where $N_p$ is the number of particles in each configuration, $N_c$ is the number of configurations in the ensemble, and $\rho = N_p/L^2$ is the number density in a square periodic box of side length $L$. Thus,

$$L = \sqrt{\frac{N_p}{\rho}}, \qquad k_{\min} = \frac{2\pi}{L} \tag{9}$$

Lengths are often reported in units of the ensemble-averaged nearest-neighbor distance $a$. The point configurations are used directly for the structure-factor and number variance analyses. For the spreadability method, they are decorated by identical disks, with phase 2 defined as the union of the disk domains, and the resulting structures are treated as two-phase media, as described in Sec. IV C.

The choice $N_p = 200$ is deliberate. It is not meant to approximate the thermodynamic limit. Rather, it creates a representative finite-size setting in which the relevant limitations are visible. At this size, the smallest accessible wavenumber is not very close to zero, the number of usable low-$k$ points is limited, the number-variance window must remain below a fraction of the box size, and the spreadability method has only a finite long-time interval before diffusion becomes affected by the box scale. These features make the benchmark suitable for testing finite-size window selection and method consistency. In selected tests, $N_p$ is varied to examine how the accessible low-$k$ resolution, fitting-window stability, and joint estimates change with system size.

Because the configurations are generated under a prescribed $S(k)$ constraint, the structure-factor method is more directly connected to the construction procedure than the number-variance and spreadability methods. We therefore use this dataset as a controlled methodology benchmark for finite-size window selection, method complementarity, and joint-estimator stability. The reported exponents are finite-size effective estimates under the stated protocol.

## III. FINITE-SIZE LIMITATIONS OF THE THREE METHODS

The benchmark systems introduced in Sec. II allow us to examine the main finite-size limitations of the three methods before introducing the extraction protocol. Although the structure-factor, number-variance, and spreadability methods are theoretically connected in the asymptotic regime, each method requires a different limiting operation. The structure-factor method relies on the $k \to 0$ limit, the number-variance method relies on the large-$R$ limit, and the spreadability method relies on a long-time regime in which the decay is governed by the low-$k$ spectral behavior. Recent work has shown that the measurement of small-$k$ structure factors is sensitive to spectral leakage and windowing effects, which can significantly bias the extracted hyperuniform scaling exponent [55]. This motivates the use of a regularized low-$k$ fitting window rather than a fixed cutoff procedure. In finite samples, none of these limits can be accessed directly. The practical problem is therefore not to transfer the thermodynamic-limit formulas mechanically to finite data, but to identify what each method can reliably provide within the accessible finite window.

For the structure-factor method, the first difficulty is the discreteness of the low-$k$ data. In a finite periodic box of side length $L$, the smallest nonzero wavenumber is $k_{\mathrm{min}} = 2\pi/L$. The same scale also controls the natural spacing of the reciprocal-space grid. In the diagnostic tests below, the radial bin width is taken as $k_{\mathrm{bin}} = 2\pi/L$ [48-50]. Thus, when the system size decreases, the number of available low-$k$ points decreases rapidly. This makes a direct log-log fit of $S(k) \sim k^{\alpha}$ increasingly sensitive to the chosen fitting window.

To quantify the quality of a log-log fit in a candidate window, we use

$$\mathrm{RMSE}_{\log} = \left[\frac{1}{n}\sum_{i=1}^{n}\left(\log S(k_i) - \log \hat{S}(k_i)\right)^2\right]^{\frac{1}{2}} \tag{10}$$

where $n$ is the number of fitted points and $\hat{S}(k_i)$ is the value predicted by the linear fit in log-log coordinates. A small $\mathrm{RMSE}_{\log}$ means that the selected data points are close to a local power law. It does not, by itself, guarantee that the selected window corresponds to the desired low-$k$ asymptotic regime.

This point is illustrated by the target system with $\alpha_{\mathrm{theory}} = 3.0$. We fix $\rho = 1.0$, $N_c = 100$, and $k_{\mathrm{bin}} = 2\pi/L$, and fit the ensemble-averaged $S(k)$ over the empirical window $ka \in [0,2.5]$, where $a$ is the ensemble-averaged nearest-neighbor distance.

Table I shows that, as $N_p$ decreases from 10000 to 100, the number of fitted low-$k$ points decreases from 56 to 5. At the same time, the extracted exponent shifts from 2.989 to 3.275. The fitting error remains of the same order. Thus, a visually acceptable or low-error fit may still yield a finite-size effective exponent that is systematically displaced from the target value.

**TABLE I.** Structure-factor fits under a fixed empirical window for different $N_p$. All systems use $\rho = 1.0$, $N_c = 100$, $k_{\text{bin}} = 2\pi/L$, and the fitting window $ka \in [0,2.5]$. Here $N_p$ is the number of particles, $L$ is the box length, $k_{\min}$ is the smallest nonzero wavenumber, $N_{\text{fit}}$ is the number of fitted points, $\alpha$ is the extracted exponent, $\Delta\alpha = |\alpha - \alpha_{\text{theory}}|$, and $\text{RMSE}_{\log}$ is the log-log fitting error.

| $N_p$ | $L$ | $k_{\min}$ | $N_{\text{fit}}$ | $\alpha$ | $\Delta\alpha$ | $\text{RMSE}_{\log}$ |
|---|---|---|---|---|---|---|
| 10000 | 100.00 | 0.063 | 56 | 2.989 | 0.011 | 0.0351 |
| 5000 | 70.71 | 0.089 | 39 | 3.160 | 0.160 | 0.0558 |
| 1000 | 31.62 | 0.199 | 17 | 3.189 | 0.189 | 0.0427 |
| 500 | 22.36 | 0.281 | 12 | 3.201 | 0.201 | 0.0414 |
| 200 | 14.14 | 0.444 | 7 | 3.273 | 0.273 | 0.0390 |
| 100 | 10.00 | 0.628 | 5 | 3.275 | 0.275 | 0.0415 |

**TABLE II.** Structure-factor fits for the representative system with $N_p = 200$ and $\alpha_{\text{theory}} = 3.0$, obtained by fixing the left endpoint and increasing the right endpoint. The table illustrates the sensitivity of the extracted exponent to the fitting-window boundary.

| $N_{\text{fit}}$ | $ka$ range | $\alpha$ | $\Delta\alpha$ | $\text{RMSE}_{\log}$ |
|---|---|---|---|---|
| 5 | [0.461, 1.692] | 3.323 | 0.323 | 0.0432 |
| 6 | [0.461, 2.000] | 3.302 | 0.302 | 0.0401 |
| 7 | [0.461, 2.307] | 3.273 | 0.273 | 0.0389 |
| 8 | [0.461, 2.615] | 3.262 | 0.262 | 0.0368 |
| 9 | [0.461, 2.922] | 3.247 | 0.247 | 0.0357 |
| 10 | [0.461, 3.230] | 3.251 | 0.251 | 0.0339 |
| 11 | [0.461, 3.538] | 3.244 | 0.244 | 0.0326 |
| 12 | [0.461, 3.845] | 3.194 | 0.194 | 0.0470 |

A second difficulty in the $S(k)$ method is the sensitivity of the fitted exponent to the right and left boundaries of the window. Table II and Figure 1 show a representative example for $N_p = 200$ and $\alpha_{\text{theory}} = 3.0$. The ensemble-averaged $S(k)$ has a short rising low-$ka$ region, followed by a crossover toward a broader plateau. When the left endpoint is fixed and the right endpoint is moved outward, the extracted exponent and $\text{RMSE}_{\log}$ both change. As $N_{\text{fit}}$ increases from 5 to 11, $\alpha$ decreases from 3.323 to 3.244. When the twelfth point is included, the window reaches further into the crossover region, and $\alpha$ decreases to 3.194 while the fitting error increases. This behavior shows that a fixed empirical window is not sufficient. A finite-size protocol must evaluate both local log-log linearity and sensitivity to small boundary perturbations.

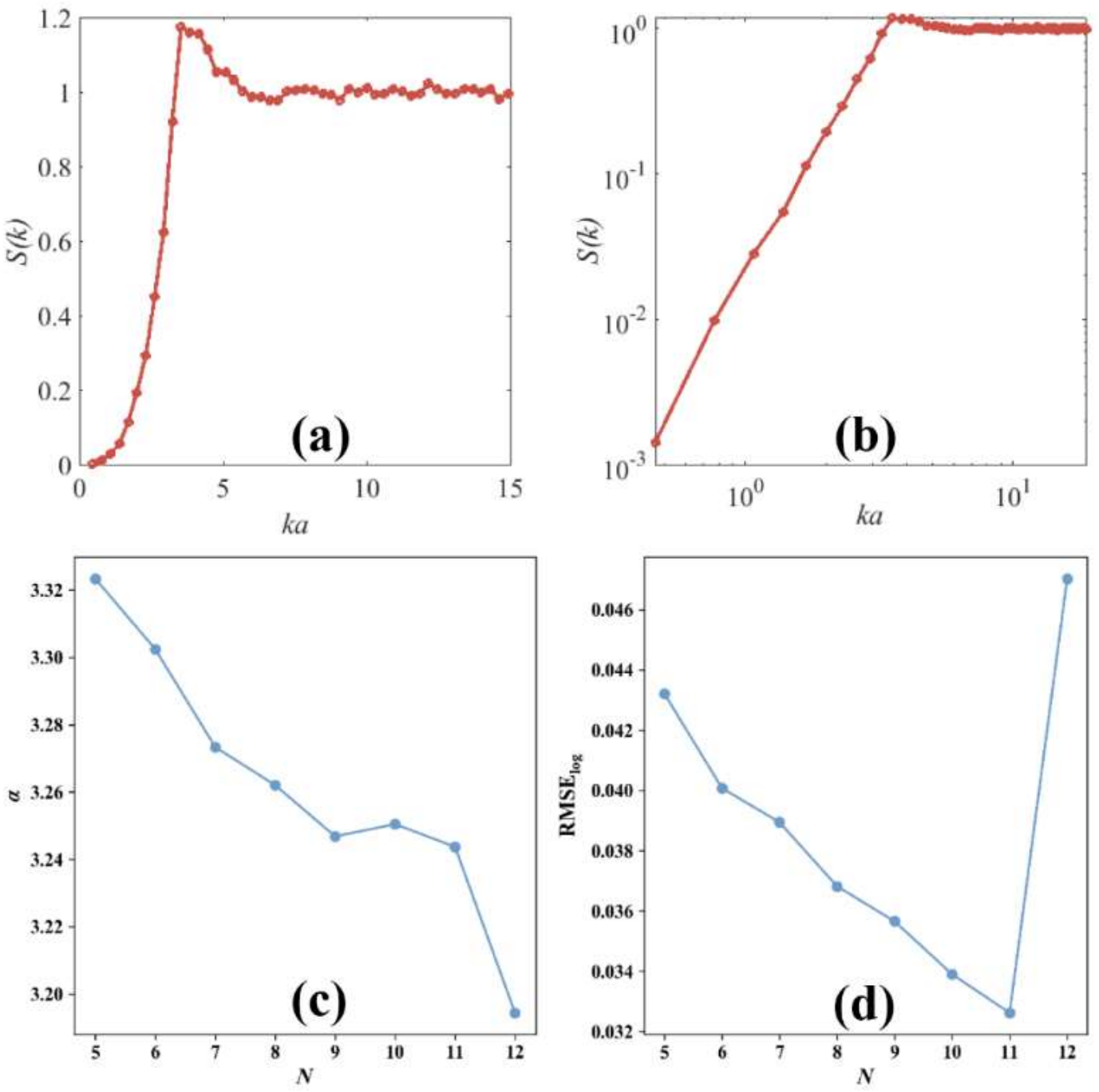

**FIG. 1.** Sensitivity of the ensemble-averaged $S(k)$ fit to the fitting window for the representative system with $N_p = 200$ and $\alpha_{\text{theory}} = 3.0$. Panel (a) shows $S(k)$ as a function of $ka$ on a linear scale. Panel (b) shows the same data on a log-log scale. Panel (c) shows the extracted exponent $\alpha$ as a function of $N_{\text{fit}}$. Panel (d) shows $\text{RMSE}_{\log}$ as a function of $N_{\text{fit}}$.

The number-variance method faces a different limitation. Its main difficulty is not only window selection, but also the capability boundary of the method itself. In two dimensions, the large-$R$ behavior of $\sigma_N^2(R)$ carries direct exponent information in the Class III regime, where $\sigma_N^2(R) \sim R^{2-\alpha}$. In the Class II and Class I regimes, however, the leading behaviors become $R\ln R$ and $R$, respectively. In these cases, number-variance remains useful for Class-like diagnosis, but it becomes less suitable as a general numerical inversion method for $\alpha$ [1, 3-4, 46].

Finite size further amplifies this limitation. In practice, the observation window must remain well inside the finite simulation domain [56]. We therefore restrict the number-variance analysis to $R < L/4$, following the usual caution that larger windows are strongly affected by finite-size and boundary effects. At small $R$, $\sigma_N^2(R)$ is still influenced by local packing and short-range structure. At large $R$, the accessible range is limited by the box size. The stable tail region that can represent large-scale behavior is therefore often short. For this reason, the number-variance method must first identify whether a tail platform is present, and then determine whether the method can provide exponent information, Class-like information, or only a qualitative trend.

The spreadability method has a third type of finite-size limitation. Compared with the direct low-$k$ fit of $S(k)$, the spreadability excess $S(\infty) - S(t)$ is often smoother because it integrates spectral information through a Gaussian diffusion kernel [14-17]. This makes it an attractive complementary method for finite systems. Nevertheless, a smoother curve does not automatically make exponent extraction reliable. The central issue is shifted from selecting a low-$k$ window to selecting a long-time fitting window. If the window is too early, the decay may still include preasymptotic contributions from higher-$k$ structure. If the window is too late, the diffusion length, of order $\sqrt{Dt}$, may become comparable to the box size, and finite-size or periodic-boundary effects can distort the decay. Thus, the spreadability method requires a finite-size-aware long-time window protocol rather than a purely visual selection of a straight segment.

Taken together, the three methods face different finite-size limitations, but they point to the same methodological need. The $S(k)$ method requires a reproducible low-$k$ window selection rule. The number-variance method requires a tail-platform analysis together with a capability-boundary check. The spreadability method requires a long-time window protocol that avoids both early preasymptotic behavior and late finite-size contamination. Therefore, effective exponent extraction in finite-size hyperuniform systems is not a single-method fitting problem. It is a combined problem of window identification, method capability assessment, and cross-method consistency. The next section introduces the corresponding protocols for the three methods.

# IV. EFFECTIVE-EXPONENT EXTRACTION PROTOCOL FOR FINITE-SIZE SYSTEMS

To extract effective scaling exponents from finite-size hyperuniform systems, we organize the three methods discussed above into a common finite-size workflow, as summarized in Figure 2. Starting from the same benchmark finite-size configurations, the workflow first computes three structural observables: the static structure factor $S(k)$, the number variance $\sigma_N^2(R)$, and the spreadability excess $S(\infty) - S(t)$, which probe reciprocal-space, real-space, and dynamic large-scale structural information, respectively [1, 3-4, 14-17, 46]. These observables are then analyzed by method-specific procedures that reflect their different finite-size roles. The $S(k)$ method uses regularized low-$k$ window screening to obtain the reciprocal-space estimate $\alpha_k$. The number-variance method uses tail-platform identification for Class-like diagnosis and provides $\alpha_{NV}$ only when the finite-size data are diagnosed as Class III-like. The spreadability method uses a plateau-window protocol to select a shared long-time fitting window and obtain the dynamic estimate $\alpha_t$. The available method outputs are finally compared and conditionally combined through a joint empirical estimator, with $u_{joint}$ reporting the internal dispersion among the participating method estimates.

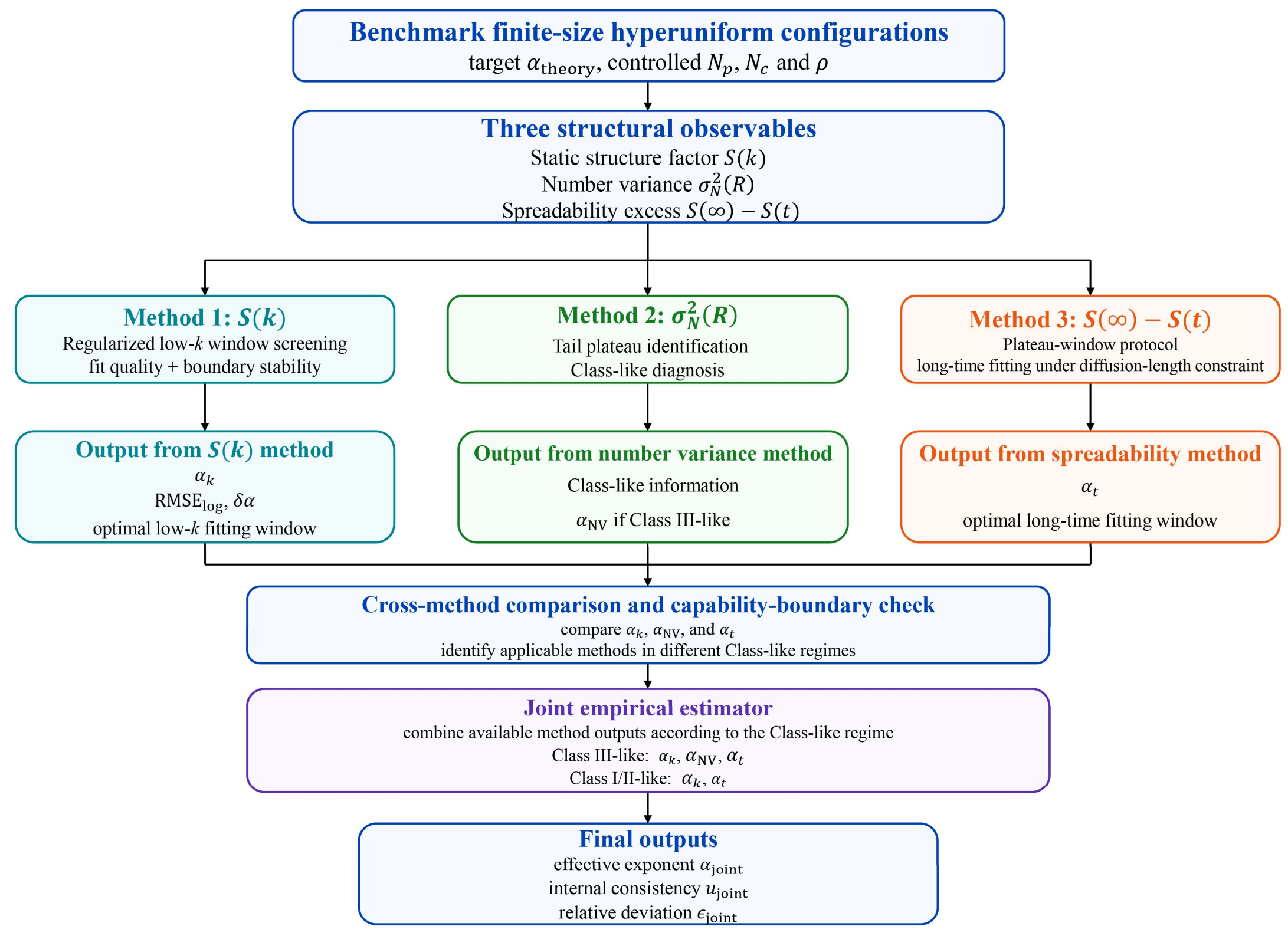

**FIG. 2.** Schematic workflow for extracting effective scaling exponents in finite-size hyperuniform systems. Starting from benchmark finite-size configurations with prescribed $\alpha_{\text{theory}}$, the protocol computes $S(k)$, $\sigma_N^2(R)$, and $S(\infty)-S(t)$. The $S(k)$ method yields $\alpha_k$ through regularized low-$k$ window screening. The number-variance method provides Class-like diagnosis and gives $\alpha_{\text{NV}}$ only in Class III-like cases. The spreadability method yields $\alpha_t$ through the plateau-window protocol. The available method outputs are then combined to obtain the joint effective estimate $\alpha_{\text{joint}}$, the internal consistency $u_{\text{joint}}$, and the relative deviation $\epsilon_{\text{joint}}$.

The three method-specific branches in Figure 2 correspond to Secs. IV A–C, and Sec. IV D defines the joint empirical estimator used to combine the method-specific outputs.

### A. Regularized low-$k$ window screening for the $S(k)$ method

The finite-size difficulty of the structure-factor method is not only that low-$k$ data points are limited. It is also that the left and right boundaries of the fitting window are not uniquely determined. If the right boundary is too large, crossover or plateau data may enter the fitting window. If the leftmost points are still affected by preasymptotic behavior, forcing the fit to start from the smallest accessible $k$ may also bias the result. We therefore replace a fixed empirical window by a regularized window-screening procedure, motivated by the sensitivity of finite-$S(k)$ analyses to window selection and reciprocal-space binning [48, 57].

We analyze the shell- and ensemble-averaged structure-factor $S(k)$. The dimensionless wavenumber is $x = ka$, where $a$ is the ensemble-averaged nearest-neighbor distance. For a contiguous candidate window $W_{ij} = [x_i, x_j]$, we fit $\log S(k) = \alpha \log(ka) + b$ in log-log coordinates. This gives the window-specific slope $\alpha_{ij}$, intercept $b_{ij}$, and fitting error

$$\text{RMSE}_{\log}(i,j) = \left[\frac{1}{n_{ij}} \sum_{m=i}^{j} \left(\log S(k_m) - \log \hat{S}(k_m)\right)^2\right]^{\frac{1}{2}} \tag{11}$$

where $n_{ij} = j - i + 1$, and $\hat{S}(k_m)$ is the value predicted by the linear fit. This error measures local log-log linearity, but it does not measure the sensitivity of the exponent to the window boundary.

To quantify boundary stability, we perturb the candidate window by moving one boundary point inward. The left- and right-boundary sensitivities are defined as

$$\delta\alpha_{ij}^{(L)} = \left|\alpha_{ij} - \alpha_{i+1,j}\right|, \quad \delta\alpha_{ij}^{(R)} = \left|\alpha_{ij} - \alpha_{i,j-1}\right| \tag{12}$$

The boundary sensitivity of the window is then

$$\delta\alpha_{ij} = \max\left(\delta\alpha_{ij}^{(L)}, \delta\alpha_{ij}^{(R)}\right) \tag{13}$$

A small $\delta\alpha_{ij}$ indicates that the fitted exponent is stable under small boundary perturbations. A large value means that the result still depends strongly on the precise choice of the fitting window.

We scan contiguous candidate windows within the low-$k$ branch of the ensemble-averaged structure factor. The admissible range is restricted to wavenumbers before the first principal peak, or before the crossover toward the broad $S(k) \sim O(1)$ plateau when no sharp peak is present. This pre-peak restriction is imposed because the exponent $\alpha$ characterizes the $k \to 0$ scaling, whereas post-peak or plateau regions mainly reflect intermediate- and short-range structure. Within this admissible low-$k$ range, candidate windows are required to contain at least $N_{\min}$ points and to span at least $\Delta_{\min}$ in $\log_{10}(ka)$.

To combine log-log fitting quality and boundary stability, we use the normalized score

$$Q_{ij} = \frac{\mathrm{RMSE}_{\log}(i,j)}{R_0} + \eta \frac{\delta\alpha_{ij}}{D_0} \tag{14}$$

where $R_0$ is the median of $\mathrm{RMSE}_{\log}$ over all candidate windows, $D_0$ is the median of $\delta\alpha$, and $\eta$ controls the relative weight of boundary stability. We use $\eta = 1.0$ as the default value, which gives equal weight to fitting quality and boundary stability after normalization. The candidate window with the smallest $Q_{ij}$ is selected as the optimal low-$k$ fitting window.

The output of the $S(k)$ method is the fitted exponent $\alpha_k = \alpha_{ij}^{*}$, together with the selected window, $N_{\mathrm{fit}}$, $\mathrm{RMSE}_{\log}$, $\delta\alpha$, and $Q$. This procedure does not aim to find the visually straightest segment. It aims to identify a finite-size effective window that is both locally power-law-like and stable with respect to small boundary changes. Sensitivity tests for $N_{\min}$, $\Delta_{\min}$, and $\eta$ are provided in the supplementary material. The results show that, near the default parameter set, the main outputs of the regularized window screening remain stable.

## B. Tail-platform identification and Class-like diagnosis from number-variance

For the number-variance method, we do not attempt to construct a universal exponent inversion procedure for all hyperuniformity classes. This choice follows from the finite-size capability of the method. In two dimensions, the Class III-like regime retains exponent information through $\sigma_N^2(R) \sim R^{2-\alpha}$, whereas the leading behaviors in the Class II-like and Class I-like regimes are $R\ln R$ and $R$, respectively [1, 3-4, 46]. Therefore, number-variance is used primarily as a Class-like diagnostic method. It contributes a numerical exponent only when the tail behavior indicates a Class III-like regime.

We characterize the large-$R$ behavior by a local effective exponent. For a local window centered around $R$, we fit $\log\sigma_N^2(R) = p\log R + c$ over $w_{\mathrm{local}}$ neighboring points. The resulting local slope is denoted by $p_{\mathrm{eff}}(R)$.

Unless otherwise stated, we use $w_{\mathrm{local}} = 5$. The next step is to identify a stable tail platform in $p_{\mathrm{eff}}(R)$. A candidate platform must satisfy two basic constraints. First, it must contain at least $M_{\min}$ points. Second, it must span at least $\Delta_{\mathrm{NV},min}$ in the $R/a$ coordinate. The default values are $M_{\min} = 5$, $\Delta_{\mathrm{NV},min} = 0.05$.

For a candidate platform $P$ containing $M$ points, we compute the mean local exponent

$$\bar{p}_{\mathrm{eff}} = \frac{1}{M} \sum_{R_m \in P} p_{\mathrm{eff}}(R_m) \tag{15}$$

together with the standard deviation $\mathrm{std}(p_{\mathrm{eff}})$ and the mean local fitting error $\mathrm{RMSE}_{\mathrm{local}}$. The latter is obtained from the local log-log fits used to evaluate $p_{\mathrm{eff}}(R)$. A stable platform should have both small local-exponent variation and good local fitting quality.

To rank candidate platforms, we use the normalized score

$$Q_{\mathrm{NV}} = \frac{\mathrm{std}(p_{\mathrm{eff}})}{P_0} + \eta_{\mathrm{NV}} \frac{\mathrm{RMSE}_{\mathrm{local}}}{E_0} \tag{16}$$

where $P_0$ and $E_0$ are the median values of $\mathrm{std}(p_{\mathrm{eff}})$ and $\mathrm{RMSE}_{\mathrm{local}}$ over all candidate platforms, respectively. The parameter $\eta_{\mathrm{NV}}$ controls the relative weight of local fitting quality. Unless otherwise stated, we use $\eta_{\mathrm{NV}} = 1.0$. The candidate platform with the smallest $Q_{\mathrm{NV}}$ is selected as the tail platform for the number-variance method.

After the platform is selected, we assign the Class-like diagnosis according to $\bar{p}_{\mathrm{eff}}$. In the present finite-size protocol, $\bar{p}_{\mathrm{eff}} \leq 1.05$ is identified as Class I-like behavior, $1.05 < \bar{p}_{\mathrm{eff}} < 1.15$ is identified as Class II-like behavior, and $\bar{p}_{\mathrm{eff}} \geq 1.15$ is identified as Class III-like behavior. These thresholds are not intended to redefine the thermodynamic-limit hyperuniformity classes. They are operational thresholds for finite-size Class-like diagnosis under the present protocol.

Only in the Class III-like case do we convert the number-variance platform into an approximate exponent estimate,

$$\alpha_{\mathrm{NV}} = 2 - \bar{p}_{\mathrm{eff}} \tag{17}$$

For Class I-like and Class II-like cases, no numerical $\alpha_{\mathrm{NV}}$ is assigned. In those cases, the number-variance method contributes only diagnostic information. This treatment avoids forcing an ill-conditioned exponent inversion onto number-variance data in regimes where the leading growth no longer distinguishes different values of $\alpha > 1$.

The output of the number-variance method therefore includes the selected tail platform, $\bar{p}_{\mathrm{eff}}$, $\mathrm{std}(p_{\mathrm{eff}})$, $\mathrm{RMSE}_{\mathrm{local}}$, the Class-like diagnosis, and, only when applicable, $\alpha_{\mathrm{NV}}$. Sensitivity tests for $w_{\mathrm{local}}$, $M_{\min}$, $\Delta_{\mathrm{NV},min}$, and $\eta_{\mathrm{NV}}$ are provided in the supplementary material. The results show that, near the default parameter set, the selected tail platform and the resulting Class-like diagnosis remain stable. For the representative system with $\alpha_{\mathrm{theory}} = 3.0$, the number-variance method consistently gives a near-linear Class I-like diagnosis, supporting its role as a diagnostic method rather than a universal exponent-inversion method.

### C. Plateau-window protocol for spreadability excess $S(\infty) - S(t)$

For the spreadability method, each point configuration is first decorated as a two-phase medium. We define phase 2 as the union of identical disks centered at the particle positions, and phase 1 as the complement in the periodic simulation box. The disk radius is denoted by $R_d$. In the present benchmark, the nominal decoration fraction is fixed at $\phi_2 = 0.005$, and $R_d$ is set by the dilute-decoration relation $\phi_2 = \rho\pi R_d^2$ with $\rho = 1.0$. Because phase 2 is defined as the union of disks, this nominal construction does not impose a strict non-overlap constraint. The relevant requirement for the spreadability method is instead that the decorated two-phase medium preserves the low-$k$ spectral exponent that controls the long-time decay. In the supplementary material, we verify this point by directly computing the spectral density $\tilde{\chi}_V(k)$ of the actual disk-union media for several decoration fractions. These tests show that $\phi_2 = 0.005$ preserves the finite-window low-$k$ exponent despite residual disk overlaps, while larger decorations such as $\phi_2 = 0.05$ produce stronger finite-radius and overlap corrections. Therefore, $\phi_2 = 0.005$ is used as a practical default that balances dilute-decoration behavior, exponent preservation, numerical stability, and computational cost. The spreadability calculation is then performed directly on this decorated two-phase medium.

Numerically, $S(t)$ is obtained by solving the diffusion equation on the periodic disk-union medium in Fourier space. We initialize a concentration field with unit concentration in phase 2 and zero concentration in phase 1,

$$c(\mathbf{x},0) = \mathcal{I}_2(\mathbf{x}) \tag{18}$$

where $\mathcal{I}_2(\mathbf{x})$ is the phase-2 indicator function. The concentration field evolves according to

$$\frac{\partial c(\mathbf{x},t)}{\partial t} = D\nabla^2 c(\mathbf{x},t) \tag{19}$$

Under periodic boundary conditions, this equation is solved by fast Fourier transforms. In Fourier space, the solution is

$$\hat{c}(\mathbf{k},t) = \hat{c}(\mathbf{k},0)\exp(-D|\mathbf{k}|^2 t) \tag{20}$$

After inverse Fourier transformation, the spreadability is computed as the fraction of the initially phase-2 mass that has entered phase 1,

$$S(t) = \frac{1}{\phi_{2,\text{act}}V}\int_V \mathcal{I}_1(\mathbf{x})c(\mathbf{x},t)\,d\mathbf{x} \tag{21}$$

where $\mathcal{I}_1(\mathbf{x}) = 1 - \mathcal{I}_2(\mathbf{x})$, and $\phi_{2,\text{act}}$ is the actual phase-2 volume fraction of the disk-union medium. With this convention, the long-time limit is $S(\infty) = \phi_1$.

The diffusion coefficient is fixed at $D = 1$. We introduce the dimensionless time $\tau = Dt/R_d^2$. For compact notation, we write the spreadability excess as $E(\tau) = S(\infty) - S(t)$. This nondimensionalization connects the diffusion time to the characteristic length scale of the disk-decorated two-phase medium and allows different target systems to be compared using the same time variable.

In the asymptotic long-time regime, the spreadability excess is connected to the small-$k$ scaling of the spectral density [15-17, 47]. In two dimensions, if the relevant low-$k$ exponent is $\alpha$, then

$$E(\tau) \sim \tau^{-\frac{2+\alpha}{2}} \tag{22}$$

Therefore, if a log-log fit of $E(\tau)$ against $\tau$ over a given time window gives the slope $m$, the corresponding exponent is

$$\alpha = -2m - 2 \tag{23}$$

The difficulty is that, in a finite system, the appropriate long-time fitting window is not known in advance. A window chosen too early may still contain preasymptotic contributions from higher-$k$ structure. A window chosen too late may be affected by the finite box size and periodic-boundary effects. We therefore use a plateau-window protocol to identify a reproducible fitting interval on the ensemble-averaged $E(\tau)$ curve.

The protocol first restricts candidate windows to a finite long-time search region guided by the diffusion length. The effective wavenumber scale probed at time $\tau$ can be estimated, at the order-of-magnitude level, as

$$k_{\text{eff}}(\tau) \sim \frac{1}{R_d\sqrt{\tau}} \tag{24}$$

Thus, a useful fitting window should lie in an intermediate regime. It should be late enough for the decay to be dominated by low-$k$ information, but not so late that $k_{\text{eff}}$ becomes comparable to the smallest accessible wavenumber of the finite system. The theoretical interpretation of this finite-size admissible window is given in the Appendix. In the actual protocol, this condition is used as a diffusion-length constraint to remove clearly too-early and too-late candidate windows.

For each remaining candidate time window $W_{ij}^{(t)} = [\tau_i, \tau_j]$, we perform a linear fit of $\log E(\tau) = m\log\tau + c$ and compute the fitting error

$$\text{RMSE}_{\log}^{(t)}(i,j) = \left[\frac{1}{n_{ij}^{(t)}}\sum_{m=i}^{j}\left(\log E(\tau_m) - \log\hat{E}(\tau_m)\right)^2\right]^{\frac{1}{2}} \tag{25}$$

where $n_{ij}^{(t)}$ is the number of fitted time points, and $\hat{E}(\tau_m)$ is the value predicted by the linear fit. To avoid very narrow fitting intervals, each candidate window must satisfy $n_{ij}^{(t)} \geq N_{t,\min}$ and $\log_{10}(\tau_j) - \log_{10}(\tau_i) \geq \Delta_{t,\min}$. Unless otherwise stated, we use $N_{t,\min} = 25$ and $\Delta_{t,\min} = 0.25$.

To quantify local exponent stability, we compute a local effective exponent on the ensemble-averaged curve. Using a sliding window of $w_t$ neighboring time points, we obtain the local slope $m_{\text{eff}}(\tau)$ from a local log-log fit and define

$$\alpha_{\text{eff}}(\tau) = -2m_{\text{eff}}(\tau) - 2 \tag{26}$$

Unless otherwise stated, we use $w_t = 17$. For each candidate time window, we compute the standard deviation $\text{std}(\alpha_{\text{eff}})$ within that window. A stable plateau window should have both a small log-log fitting error and a small local-exponent variation.

We rank candidate windows using the normalized score

$$Q_t = \frac{\text{RMSE}_{\log}^{(t)}}{R_t} + \eta_t\frac{\text{std}(\alpha_{\text{eff}})}{A_t} \tag{27}$$

where $R_t$ and $A_t$ are the median values of $\text{RMSE}_{\log}^{(t)}$ and $\text{std}(\alpha_{\text{eff}})$ over all candidate time windows. The parameter $\eta_t$ controls the relative weight of local exponent stability. Unless otherwise stated, we use $\eta_t = 1.0$. The candidate window with the smallest $Q_t$ is selected as the plateau window.

The selected plateau window is determined from the ensemble-averaged $E(\tau)$ curve and is then applied as a shared fitting window to all configurations in the same ensemble. This shared-window strategy avoids introducing additional fluctuations from configuration-dependent window choices. The output of the spreadability method is the exponent $\alpha_t$, together with the selected time window, $\mathrm{RMSE}_{\log}^{(t)}$, $\mathrm{std}(\alpha_{\mathrm{eff}})$, and $Q_t$.

Sensitivity tests for $w_t$, $\Delta_{t,\min}$, $N_{t,\min}$, and $\eta_t$ are provided in the supplementary material. The results show that, near the default parameter set, the selected plateau window and the extracted spreadability exponent remain stable. Thus, the plateau-window protocol regularizes the finite long-time fitting problem without changing the theoretical relation between the spreadability excess and the scaling exponent.

### D. Joint empirical estimator and internal consistency

The three method-specific outputs have different meanings. The $S(k)$ method provides a direct reciprocal-space estimate $\alpha_k$. The spreadability method provides a dynamic estimate $\alpha_t$ from the long-time decay of $S(\infty) - S(t)$. The number-variance method provides Class-like diagnostic information and gives a numerical estimate $\alpha_{\mathrm{NV}}$ only when the selected tail platform is identified as Class III-like. We therefore combine the method outputs conditionally rather than forcing all three methods to contribute in every case. The joint quantity defined below is not intended to be a statistically optimal estimator. It is used as a transparent finite-size consensus summary of the method estimates available under the stated protocol.

The joint empirical estimator is defined as

$$\alpha_{\mathrm{joint}} = \frac{\alpha_k + \alpha_t + \lambda_{\mathrm{III}}\alpha_{\mathrm{NV}}}{2 + \lambda_{\mathrm{III}}} \tag{28}$$

Here, $\lambda_{\mathrm{III}}$ is an availability indicator for the number-variance estimate. We set $\lambda_{\mathrm{III}} = 1$ when the number-variance method is identified as Class III-like, so that $\alpha_{\mathrm{NV}}$ enters the joint estimate as a third contribution. We set $\lambda_{\mathrm{III}} = 0$ when the number-variance method is identified as Class I-like or Class II-like, so that the joint estimate is obtained only from $\alpha_k$ and $\alpha_t$. This conditional rule reflects the capability boundary of the number-variance method. It allows number-variance to contribute when its large-$R$ behavior retains exponent information, but avoids using it as an ill-conditioned inversion method in regimes where the leading number-variance growth no longer distinguishes different values of $\alpha > 1$.

To quantify the agreement among the participating methods, we define the internal consistency as

$$u_{\text{joint}} = \left[ \frac{1}{N_{\text{method}}} \sum_{\alpha_i \in \mathcal{A}} \left( \alpha_i - \alpha_{\text{joint}} \right)^2 \right]^{\frac{1}{2}} \tag{29}$$

where $\mathcal{A}$ is the set of method estimates included in the joint estimator, and $N_{\text{method}}$ is the number of included estimates. For Class I-like and Class II-like cases, $\mathcal{A} = \{\alpha_k, \alpha_t\}$. For Class III-like cases, $\mathcal{A} = \{\alpha_k, \alpha_t, \alpha_{\text{NV}}\}$.

The quantity $u_{\text{joint}}$ measures the dispersion among the participating method estimates. It should not be interpreted as a statistical confidence interval, a standard error, or an asymptotic uncertainty of the true exponent. A small $u_{\text{joint}}$ indicates that the participating finite-size methods give mutually consistent estimates under the same protocol. A large $u_{\text{joint}}$ indicates method-level disagreement and should be treated as a warning that the effective exponent may be sensitive to window selection, method capability, or finite-size contamination.

For the controlled benchmark systems studied here, the prescribed target exponent $\alpha_{\text{theory}}$ is known. We therefore also report the absolute deviation

$$\Delta\alpha_{\text{joint}} = \left| \alpha_{\text{joint}} - \alpha_{\text{theory}} \right| \tag{30}$$

and the relative deviation

$$\epsilon_{\text{joint}} = \frac{\left| \alpha_{\text{joint}} - \alpha_{\text{theory}} \right|}{\alpha_{\text{theory}}} \tag{31}$$

These two quantities are used only for benchmark evaluation. In applications to experimental data or to systems without a prescribed target exponent, $\alpha_{\text{theory}}$ is unavailable, and $\Delta\alpha_{\text{joint}}$ and $\epsilon_{\text{joint}}$ cannot be computed. In that case, the reported outputs should be $\alpha_{\text{joint}}$, $u_{\text{joint}}$, the method-specific estimates, the selected fitting windows, and the Class-like diagnosis.

The joint empirical estimator is therefore a finite-size summary of the available method outputs under the stated analysis protocol. It provides a reproducible finite-size effective estimate for comparing systems analyzed under the same protocol, rather than a reconstruction of the thermodynamic-limit asymptotic exponent.

# V. RESULTS

In this section, we apply the finite-size extraction protocol to the benchmark target hyperuniform systems described in Section II. The results are organized according to the three method-specific analyses and the final joint estimate. Section V A presents the structure-factor method and examines how the regularized low-$k$ window screening improves the stability of $S(k)$-based exponent extraction. Section V B analyzes the number-variance method and its Class-like diagnostic capability. Section V C presents the spreadability method and the plateau-window results. Section V D combines the available method outputs through the joint empirical estimator. Section V E provides an external validation test using an independently generated Gaussian pair-statistics system.

## A. Static structure-factor $S(k)$

We first examine the performance of the regularized low-$k$ window screening for the structure-factor method. As a representative example, we consider the target system with $N_p = 200$ and $\alpha_{\text{theory}} = 3.0$. This system is useful because it exhibits two typical finite-size features. The low-$ka$ region shows a clear rising trend, but the available range is short. In addition, the extracted exponent is sensitive to the fitting-window boundary, as shown in Sec. III. This system therefore provides a stringent test of whether the proposed window-screening procedure can reduce the ambiguity of empirical window selection.

Using the regularized screening protocol, the optimal fitting window selected from the ensemble-averaged $S(k)$ is $ka \in [0.770, 2.925]$, with $N_{\text{fit}} = 8$. This window gives $\alpha_k = 3.119$, $\text{RMSE}_{\text{lo}} = 0.0174$. For comparison, the nominal fixed empirical range $ka \in [0, 2.5]$, corresponding to the available fitting points $ka \in [0.462, 2.310]$, gives $\alpha = 3.273$ and $\text{RMSE}_{\text{log}} = 0.0390$. Thus, the regularized protocol reduces the absolute deviation from 0.273 to 0.119 and simultaneously lowers the log-log fitting error. This comparison shows that a fixed empirical low-$k$ window does not necessarily provide a stable finite-size estimate, while a window selected by balancing fit quality and boundary stability gives a more stable effective low-$k$ estimate.

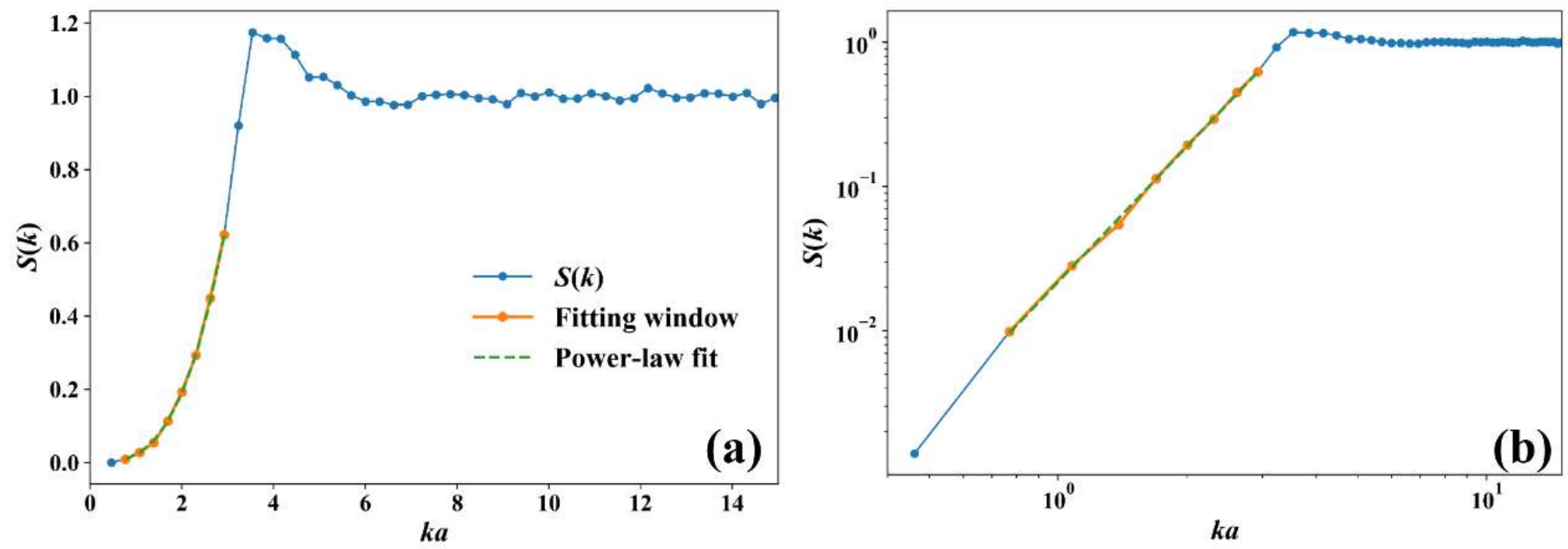

**FIG. 3.** Regularized low-$k$ fitting window for the representative system with $N_p = 200$ and $\alpha_{\text{theory}} = 3.0$. Panel (a) shows the ensemble-averaged $S(k)$ as a function of $ka$ on a linear scale. Panel (b) shows the same data on a log-log scale. The highlighted points indicate the selected fitting window, and the dashed line represents the power-law fit within that window.

Figure 3 shows the position of the selected window on the ensemble-averaged $S(k)$ curve. The selected interval lies within the rising low-$ka$ region. It avoids the leftmost points that may still be affected by preasymptotic finite-size deviations and does not extend into the post-peak crossover region. In the log-log representation, the selected points follow a nearly linear trend, indicating that the protocol does not simply choose the leftmost available data points. Instead, it selects a finite-size effective window that balances local power-law behavior and boundary stability.

**TABLE III.** Top ten candidate $S(k)$ fitting windows ranked by the normalized score $Q$ for the representative system with $N_p = 200$ and $\alpha_{\text{theory}} = 3.0$. Here $i_L$ and $i_R$ are the left and right data-point indices of the candidate window, $N_{\text{fit}}$ is the number of fitted points, $ka$ range is the fitting interval, $\alpha$ is the extracted exponent, $\Delta\alpha = |\alpha - \alpha_{\text{theory}}|$, $\text{RMSE}_{\log}$ is the log-log fitting error, $\delta\alpha$ is the boundary perturbation sensitivity, and $Q$ is the normalized score.

| $i_L$ | $i_R$ | $N_{\text{fit}}$ | $ka$ range | $\alpha$ | $\Delta\alpha$ | $\text{RMSE}_{\log}$ | $\delta\alpha$ | $Q$ |
|---|---|---|---|---|---|---|---|---|
| 2 | 9 | 8 | [0.770, 2.925] | 3.119 | 0.119 | 0.0174 | 0.0285 | 1.2598 |
| 2 | 11 | 10 | [0.770, 3.541] | 3.149 | 0.149 | 0.0183 | 0.0316 | 1.3486 |
| 2 | 8 | 7 | [0.770, 2.617] | 3.117 | 0.117 | 0.0186 | 0.0354 | 1.4076 |
| 2 | 7 | 6 | [0.770, 2.309] | 3.099 | 0.099 | 0.0192 | 0.0333 | 1.4157 |
| 2 | 10 | 9 | [0.770, 3.233] | 3.145 | 0.145 | 0.0192 | 0.0355 | 1.4429 |
| 4 | 11 | 8 | [1.386, 3.541] | 3.245 | 0.245 | 0.0133 | 0.0610 | 1.4523 |
| 4 | 10 | 7 | [1.386, 3.233] | 3.260 | 0.260 | 0.0137 | 0.0713 | 1.6029 |
| 3 | 11 | 9 | [1.078, 3.541] | 3.181 | 0.181 | 0.0176 | 0.0647 | 1.7245 |
| 2 | 6 | 5 | [0.770, 2.001] | 3.096 | 0.096 | 0.0210 | 0.0512 | 1.7332 |
| 3 | 9 | 7 | [1.078, 2.925] | 3.148 | 0.148 | 0.0176 | 0.0841 | 1.9661 |

Table III lists the ten candidate windows with the lowest values of the normalized score $Q$. The best-ranked window is $ka \in [0.770, 2.925]$, with $\alpha = 3.119$, $\mathrm{RMSE}_{\log} = 0.0174$, and $\delta\alpha = 0.0285$. Some other candidate windows have slightly smaller local fitting errors, such as the window $ka \in [1.386, 3.541]$, but they show larger boundary sensitivity or extend closer to the crossover region. This confirms that finite-size low-$k$ window selection should not be based only on minimizing the fitting error. The combined score $Q$ provides a more balanced criterion by incorporating both log-log fitting quality and boundary robustness.

We then apply the same procedure to all nine target exponents, $\alpha_{\text{theory}}$ =0.3, 0.5, 0.7, 1.0, 1.5, 2.0, 2.5, 3.0 and 4.0, and compare it with the fixed-window scheme. The results are summarized in Table IV. Across all cases, the regularized screening method reduces both the absolute deviation and the fitting error relative to the fixed-window fit. For example, at $\alpha_{\text{theory}} = 1.0$, the fixed-window estimate is 1.109, whereas the regularized estimate is 1.047. At $\alpha_{\text{theory}} = 3.0$, the fixed-window estimate is 3.273, whereas the regularized estimate is 3.119. These results show that the proposed screening procedure reduces the effect of empirical window choice and provides a more reproducible input for the subsequent cross-method analysis.

**TABLE IV.** Comparison between fixed-window and regularized $S(k)$ fitting schemes for target systems with different $\alpha_{\text{theory}}$. Here $N_{\text{fit}}$ is the number of fitted points, $ka$ range is the fitting window, $\alpha$ is the extracted exponent, $\Delta\alpha = |\alpha - \alpha_{\text{theory}}|$, and $\mathrm{RMSE}_{\log}$ is the log-log fitting error.

| $\alpha_{\text{theory}}$ | Method | $N_{\text{fit}}$ | $ka$ range | $\alpha$ | $\Delta\alpha$ | $\mathrm{RMSE}_{\log}$ |
|---|---|---|---|---|---|---|
| 0.3 | fixed-window | 10 | [0.349, 2.443] | 0.330 | 0.030 | 0.0035 |
| | regularized | 6 | [0.582, 1.745] | 0.316 | 0.016 | 0.0007 |
| 0.5 | fixed-window | 9 | [0.362, 2.294] | 0.549 | 0.049 | 0.0060 |
| | regularized | 8 | [0.604, 2.293] | 0.525 | 0.025 | 0.0012 |
| 0.7 | fixed-window | 9 | [0.373, 2.360] | 0.768 | 0.068 | 0.0081 |
| | regularized | 8 | [0.621, 2.360] | 0.735 | 0.035 | 0.0016 |
| 1.0 | fixed-window | 9 | [0.389, 2.462] | 1.109 | 0.109 | 0.0117 |
| | regularized | 8 | [0.648, 2.462] | 1.047 | 0.047 | 0.0025 |
| 1.5 | fixed-window | 8 | [0.413, 2.340] | 1.623 | 0.123 | 0.0201 |
| | regularized | 6 | [0.963, 2.340] | 1.550 | 0.050 | 0.0024 |
| 2.0 | fixed-window | 8 | [0.433, 2.451] | 2.185 | 0.185 | 0.0224 |
| | regularized | 8 | [0.721, 2.739] | 2.089 | 0.089 | 0.0054 |
| 2.5 | fixed-window | 7 | [0.448, 2.240] | 2.736 | 0.236 | 0.0286 |
| | regularized | 8 | [1.344, 3.433] | 2.680 | 0.180 | 0.0079 |
| 3.0 | fixed-window | 7 | [0.462, 2.310] | 3.273 | 0.273 | 0.0390 |
| | regularized | 8 | [0.770, 2.925] | 3.119 | 0.120 | 0.0174 |
| 4.0 | fixed-window | 7 | [0.484, 2.421] | 4.355 | 0.355 | 0.0514 |
| | regularized | 9 | [1.130, 3.712] | 4.205 | 0.205 | 0.0137 |

These results highlight the main advantage of the regularized $S(k)$ method. It keeps the exponent extraction directly tied to the reciprocal-space definition of hyperuniformity, while replacing an empirical cutoff by a reproducible window-selection rule. Across the benchmark systems, the regularized selection gives more accurate finite-size estimates than the fixed-window fit, reducing both the absolute deviation from $\alpha_{\text{theory}}$ and the log-log fitting error. Thus, the strength of the $S(k)$ method is not merely that it provides a direct exponent estimate, but that it makes the low-$k$ fitting procedure more stable and less dependent on a manually chosen cutoff.

Overall, the results show that the regularized $S(k)$ fitting scheme mitigates two finite-size difficulties at the same time: preasymptotic bias at the left end of the accessible low-$k$ region and crossover contamination near the right end of the fitting window. Compared with the fixed-window scheme, the regularized method gives a more explicit and reproducible rule for selecting the effective low-$k$ window. At the same time, the $S(k)$ method still tends to overestimate the exponent for larger $\alpha_{\text{theory}}$. This indicates that finite-size and preasymptotic effects are reduced but not eliminated. The $S(k)$ method therefore provides a clear and reproducible primary estimate, but its output should be compared with the spreadability and number-variance methods before entering the final joint estimate.

### B. Number-variance $\sigma_N^2(R)$

We next examine the number-variance method for the nine target systems with $N_p = 200$. As discussed in Sec. IV B, this method is not used as a universal exponent-inversion method. Its primary role is to provide a Class-like diagnosis from the large-$R$ behavior of $\sigma_N^2(R)$. A numerical estimate $\alpha_{\text{NV}}$ is assigned only when the selected tail platform is identified as Class III-like, following the asymptotic number-variance classification associated with $S(k) \sim k^{\alpha}$ [1, 3-4, 46].

Table V summarizes the tail-platform analysis for all nine target systems. For $\alpha_{\text{theory}} =$ 0.3, 0.5 and 0.7, the mean local exponent $\bar{p}_{\text{eff}}$ is clearly larger than the near-linear regime. These systems are therefore identified as Class III-like. In this regime, number-variance retains effective exponent information through $\sigma_N^2(R) \sim R^{2-\alpha}$, and the approximate estimate $\alpha_{\text{NV}} = 2 - \bar{p}_{\text{eff}}$ can be used. The resulting values are $\alpha_{\text{NV}} =$ 0.309, 0.492 and 0.658, respectively, which are close to the prescribed target exponents.

For $\alpha_{\text{theory}} = 1.0$, the selected tail platform gives $\bar{p}_{\text{eff}} = 1.063$, which falls into the Class II-like interval under the present operational thresholds. This is consistent with the logarithmic crossover associated with the $R\ln R$ growth. Because this is a boundary-like regime, we do not convert the number-variance output into a numerical $\alpha_{\mathrm{NV}}$.

For $\alpha_{\mathrm{theory}}$ =1.5, 2.0, 2.5, 3.0 and 4.0, the selected tail platforms are near-linear, with $\bar{p}_{\mathrm{eff}}$ close to 1. These systems are therefore identified as Class I-like. In this regime, the leading number-variance growth no longer distinguishes different values of $\alpha > 1$. Thus, the number-variance method confirms the Class I-like behavior but does not provide a numerical exponent estimate.

**TABLE V.** Number-variance tail-platform analysis for the nine target systems with $N_p = 200$. Here $M_{\mathrm{plateau}}$ is the number of points in the selected platform, $R/a$ range is the platform interval, $\bar{p}_{\mathrm{eff}}$ is the mean local effective exponent, $\mathrm{std}(p_{\mathrm{eff}})$ measures platform stability, $\mathrm{RMSE}_{\mathrm{local}}$ is the local log-log fitting error, $\sigma_N^2(R)$ class is the Class-like diagnosis from the number-variance method, and $\alpha_{\mathrm{NV}}$ is the approximate exponent estimate when available. A dash indicates that number-variance contributes only diagnostic information.

| $\alpha_{\mathrm{theory}}$ | $M_{\mathrm{plateau}}$ | $R/a$ range | $\bar{p}_{\mathrm{eff}}$ | $\mathrm{std}(p_{\mathrm{eff}})$ | $\mathrm{RMSE}_{\mathrm{local}}$ | $\sigma_N^2(R)$ class | $\alpha_{\mathrm{NV}}$ |
|---|---|---|---|---|---|---|---|
| 0.3 | 5 | [4.153, 4.707] | 1.691 | 0.0154 | $1.4\times10^{-4}$ | Class III-like | 0.309 |
| 0.5 | 5 | [3.905, 4.425] | 1.508 | 0.0083 | $1.9\times10^{-4}$ | Class III-like | 0.492 |
| 0.7 | 6 | [4.425, 5.058] | 1.342 | 0.0063 | $1.3\times10^{-4}$ | Class III-like | 0.658 |
| 1.0 | 6 | [4.600, 5.205] | 1.063 | 0.0193 | $2.8\times10^{-4}$ | Class II-like | - |
| 1.5 | 6 | [4.452, 5.023] | 1.047 | 0.0291 | $3.4\times10^{-4}$ | Class I-like | - |
| 2.0 | 7 | [3.269, 5.448] | 0.973 | 0.0298 | $3.5\times10^{-4}$ | Class I-like | - |
| 2.5 | 6 | [3.473, 3.999] | 0.955 | 0.0521 | $5.0\times10^{-4}$ | Class I-like | - |
| 3.0 | 7 | [4.286, 4.898] | 0.952 | 0.0972 | $7.9\times10^{-4}$ | Class I-like | - |
| 4.0 | 7 | [4.088, 4.672] | 0.931 | 0.1232 | $9.9\times10^{-4}$ | Class I-like | - |

Figure 4 shows three representative cases, $\alpha_{\mathrm{theory}}$=0.5, 1.0 and 3.0 corresponding to Class III-like, Class II-like, and Class I-like diagnoses. The selected platform points are highlighted on the number-variance curves. For $\alpha_{\mathrm{theory}}$ =0.5, the tail slope is clearly superlinear, and the number-variance method provides the estimate $\alpha_{\mathrm{NV}} =$ 0.492. For $\alpha_{\mathrm{theory}} = 1.0$, the tail behavior is weakly superlinear and is identified as Class II-like. For $\alpha_{\mathrm{theory}} = 3.0$, the tail is near-linear, consistent with Class I-like behavior. These three examples illustrate how the same platform-identification protocol separates the finite-size Class-like regimes without forcing numerical exponent extraction in regimes where the method has limited resolving power.

The stability indicators in Table V also support this interpretation. The selected platforms have small $\mathrm{std}(p_{\mathrm{eff}})$ and small $\mathrm{RMSE}_{\mathrm{local}}$, indicating that the platform selection is not a visual choice of a locally flat segment. It is obtained from the same rule across all target systems. The platform locations vary from one target exponent to another, but they remain within the prescribed large-$R$ search range and satisfy the minimum platform-size and span constraints.

Overall, the number-variance method should be understood as a method with a clear capability boundary. In the Class III-like regime, it provides both Class-like diagnosis and an approximate exponent reference. In the Class II-like regime, it identifies the logarithmic crossover trend but does not provide a robust numerical inversion of $\alpha$. In the Class I-like regime, it confirms the near-linear growth but cannot distinguish different $\alpha > 1$ values. For this reason, number-variance is not used as a primary exponent estimator for all classes. Instead, it provides an important structural constraint for interpreting the $S(k)$ and spreadability estimates and for deciding whether $\alpha_{\mathrm{NV}}$ should enter the joint empirical estimator.

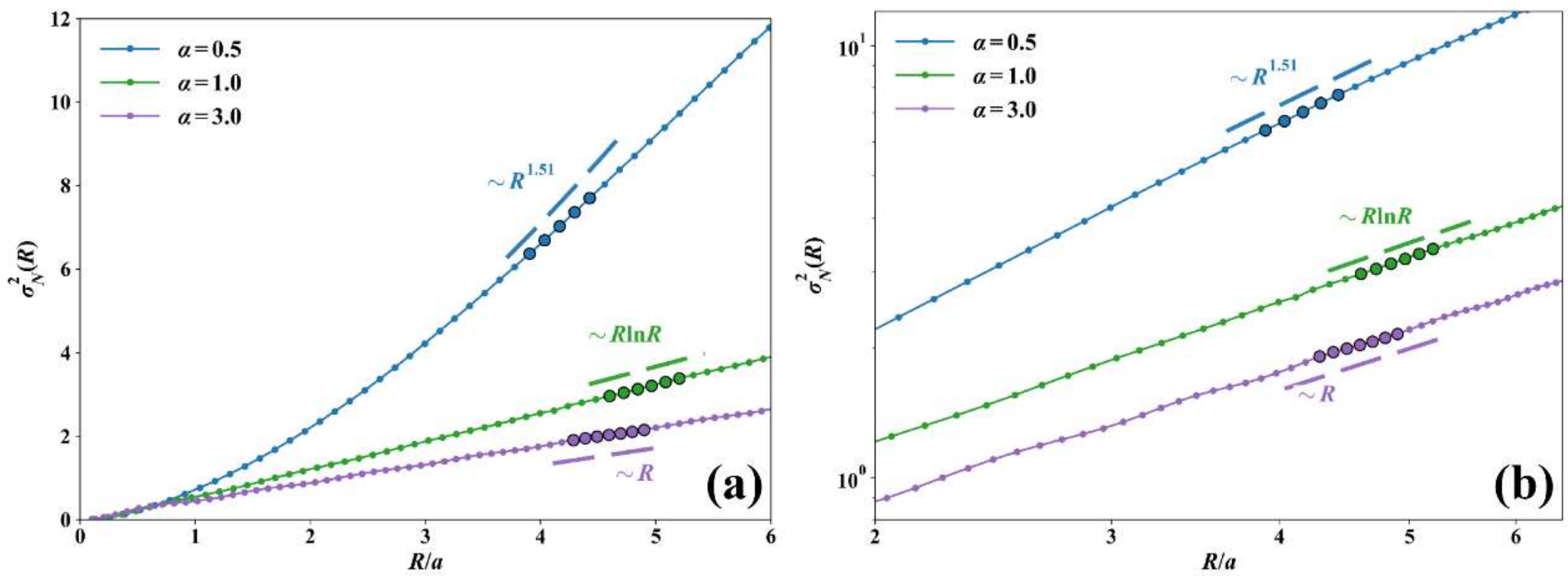

**FIG. 4.** Number-variance tail-platform identification for three representative systems. The cases $\alpha_{\mathrm{theory}}$ =0.5, 1.0 and 3.0 correspond to Class III-like, Class II-like, and Class I-like diagnoses, respectively. The markers with black edges indicate the automatically selected tail platforms. The dashed lines are guides to the eye showing the expected tail trends. The figure illustrates that number-variance provides stable Class-like diagnosis in finite samples, while its ability to provide a numerical exponent estimate depends on the diagnosed regime.

### C. Spreadability excess $S(\infty) - S(t)$

We next examine the spreadability method for the nine target systems with $N_p = 200$. As discussed in Sec. IV C, the exponent is extracted from the long-time decay of the spreadability excess $S(\infty) - S(t)$, or equivalently $E(\tau)$, where $\tau = Dt/R_d^2$ [15-17]. The main finite-size difficulty is the selection of an appropriate long-time fitting window. We therefore compare two protocols. The first is a global-baseline procedure, which performs an open search over admissible fitting windows under the diffusion-length constraint. Each candidate window must span at least one decade in $\tau$, and the selected window is the one with the smallest log-log slope variance. This baseline uses the same power-law relation but does not impose the shared-window and local-plateau stability criteria used in the plateau-window protocol. The second is the plateau-window protocol proposed in this work, which selects a shared long-time window from the ensemble-averaged curve by combining fitting quality, local effective-exponent stability, and the diffusion-length constraint.

Figure 5 shows a representative example for the target system with $\alpha_{\text{theory}} = 4.0$. The global-baseline procedure selects an early-time window, $\tau \in [0.010, 0.102]$, and gives $\alpha =$ -1.721 on the ensemble-averaged curve. This result is not a valid estimate of the long-time scaling exponent because the selected window is located in the early-time region and does not represent the decay governed by the low-$k$ spectral behavior. In contrast, the plateau-window protocol selects the shared window $\tau \in [155.022, 325.635]$, which lies in a later and more stable decay regime, and gives $\alpha_t =$ 3.928. This comparison shows that, in finite-size systems, the key issue is not only the use of a power-law relation, but also the identification of a physically admissible and reproducible time window.

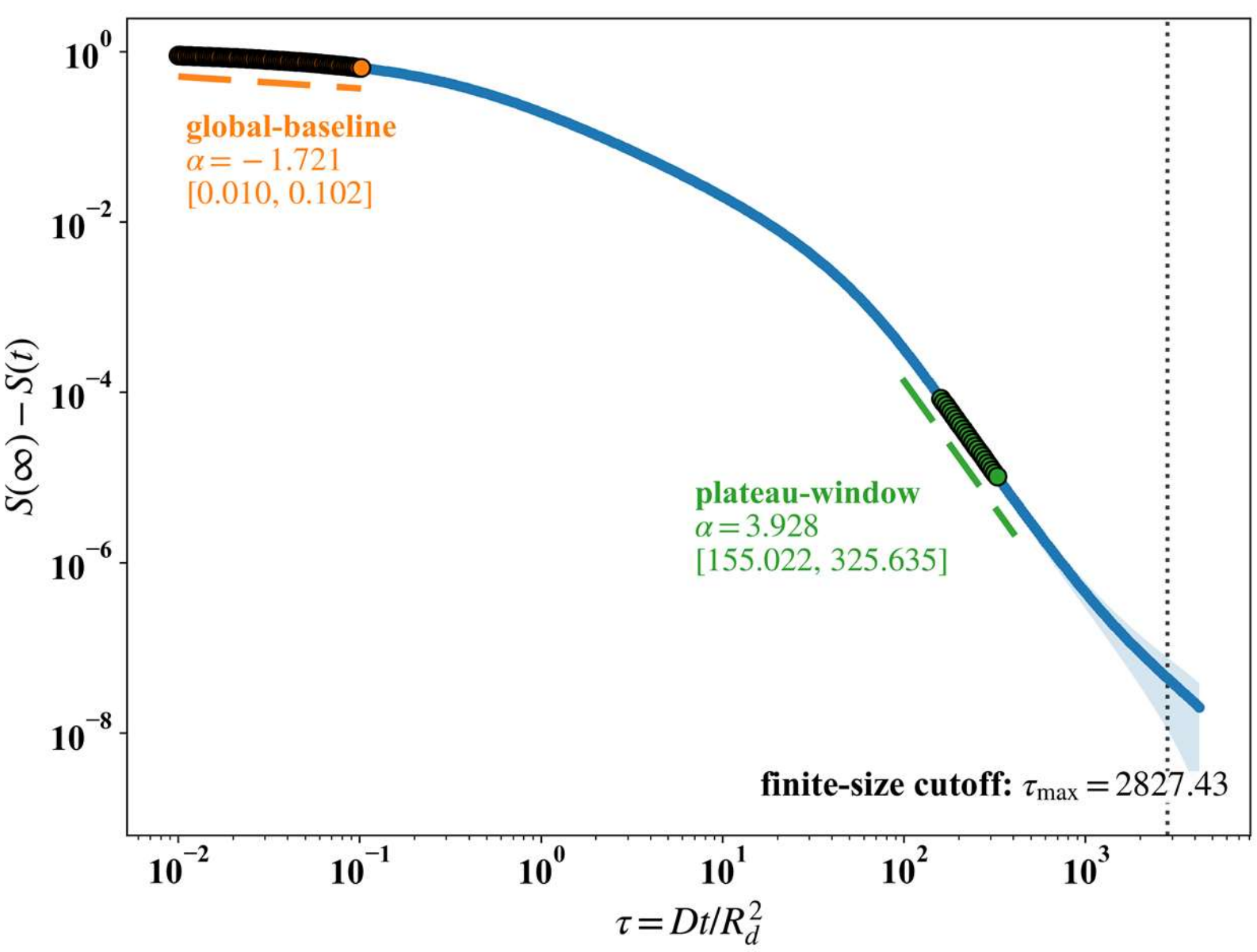

**FIG. 5.** Ensemble-averaged spreadability excess $S(\infty) - S(t)$ for the target system with $N_p =$ 200 and $\alpha_{\text{theory}} = 4.0$. The horizontal axis is the dimensionless time $\tau = Dt/R_d^2$. Markers with black edges indicate the fitting windows selected by the two protocols. The dashed line shows the corresponding fit within each selected window. The vertical dotted line indicates the finite-size diffusion-length cutoff used to delimit the admissible time range.

Table VI summarizes the results for all nine target systems. For each method, the table reports both the exponent fitted from the ensemble-averaged curve and the distribution of exponents obtained from the 100 individual configurations using the selected protocol. The global-baseline method can give reasonable ensemble-averaged estimates for some target exponents. For example, for $\alpha_{\mathrm{theory}} = 0.5$, it gives $\alpha = 0.510$ on the ensemble-averaged curve. However, the corresponding configuration-level results are widely dispersed, with $0.378 \pm 0.529$. This indicates that the open-search procedure is sensitive to configuration-level fluctuations and window drift.

The plateau-window protocol greatly reduces this configuration-to-configuration variability. For small and intermediate exponents, it gives configuration-level means close to the prescribed values while keeping the selected ensemble-level window physically admissible. The improvement is especially clear for the large-exponent stress-test case. For $\alpha_{\mathrm{theory}} = 4.0$, the global-baseline procedure selects an early-time window and gives $-0.412 \pm 2.346$ across configurations and $\alpha = -1.721$ on the ensemble-averaged curve, whereas the plateau-window protocol gives $3.952 \pm 0.275$ across configurations and $\alpha_t = 3.928$ on the ensemble-averaged curve. The latter is much closer to the prescribed target exponent and is obtained from a long-time window with smaller local-exponent variation.

The spreadability method provides a complementary advantage by converting structural information into a smooth long-time diffusion signal. The diffusion process suppresses short-scale fluctuations, while the plateau-window protocol turns the long-time fitting problem into a stability-based window-identification problem. Across the benchmark systems, the shared-window strategy reduces configuration-dependent window drift and gives more stable configuration-level exponent estimates than the open global-baseline search. Thus, the strength of the spreadability method lies in stabilizing dynamic exponent extraction under finite-size time-window constraints, rather than simply fitting a visually straight segment of $S(\infty) - S(t)$.

Overall, these results show that the plateau-window protocol improves the spreadability method in two ways. First, it selects a physically more appropriate long-time window by excluding early preasymptotic behavior and late finite-size contamination. Second, by applying a shared window to all configurations in the same ensemble, it reduces the additional variability caused by configuration-dependent window choices. The improvement is especially clear in the configuration-level standard deviations reported in Table VI.

**TABLE VI.** Comparison between the global-baseline and plateau-window protocols for extracting the spreadability exponent in the nine $N_p = 200$ target systems. The column $\bar{\alpha} \pm \mathrm{std}$ reports the mean and standard deviation of the exponents obtained from 100 configurations. $N_{\mathrm{fit}}$ and $\tau$ range denote the number of fitted time points and the selected fitting window on the ensemble-averaged curve. The column $\alpha$ reports the exponent fitted from the ensemble-averaged curve, $\Delta\alpha = |\alpha - \alpha_{\mathrm{theory}}|$, $\mathrm{RMSE}_{\log}$ is the log-log fitting error, and $\mathrm{std}(\alpha_{\mathrm{eff}})$ is the standard deviation of the local effective exponent within the selected window. For the plateau-window protocol, $\mathrm{std}(\alpha_{\mathrm{eff}})$ contributes to the window score. For the global-baseline protocol, it is evaluated after the window is selected.

| $\alpha_{\mathrm{theory}}$ | Method | $\bar{\alpha} \pm \mathrm{std}$ | $N_{fit}$ | $\tau$ range | $\alpha$ | $\Delta\alpha$ | $\mathrm{RMSE}_{\log}$ | $\mathrm{std}(\alpha_{\mathrm{eff}})$ |
|---|---|---|---|---|---|---|---|---|
| 0.3 | global-baseline | 0.218 ± 0.332 | 76 | [57.625, 586.024] | 0.317 | 0.017 | $1.70\times10^{-3}$ | $2.97\times10^{-2}$ |
| | plateau-window | 0.327 ± 0.170 | 25 | [113.785, 239.014] | 0.314 | 0.014 | $1.33\times10^{-4}$ | $6.62\times10^{-3}$ |
| 0.5 | global-baseline | 0.378 ± 0.529 | 76 | [69.374, 705.505] | 0.510 | 0.010 | $1.35\times10^{-3}$ | $2.53\times10^{-2}$ |
| | plateau-window | 0.525 ± 0.209 | 25 | [132.813, 278.983] | 0.507 | 0.007 | $8.88\times10^{-5}$ | $4.54\times10^{-3}$ |
| 0.7 | global-baseline | 0.456 ± 0.754 | 76 | [80.974, 823.482] | 0.709 | 0.009 | $1.16\times10^{-3}$ | $2.28\times10^{-2}$ |
| | plateau-window | 0.720 ± 0.189 | 25 | [150.301, 315.719] | 0.705 | 0.005 | $6.76\times10^{-5}$ | $3.53\times10^{-3}$ |
| 1.0 | global-baseline | 0.374 ± 1.163 | 76 | [132.813,1350.660] | 1.036 | 0.036 | $3.04\times10^{-4}$ | $1.12\times10^{-2}$ |
| | plateau-window | 1.058 ± 0.244 | 25 | [239.014, 502.067] | 1.039 | 0.039 | $2.38\times10^{-5}$ | $1.54\times10^{-3}$ |
| 1.5 | global-baseline | 0.464 ± 1.486 | 76 | [150.301, 1528.512] | 1.469 | 0.031 | $1.29\times10^{-4}$ | $5.29\times10^{-3}$ |
| | plateau-window | 1.520 ± 0.371 | 25 | [357.292, 750.519] | 1.471 | 0.029 | $1.98\times10^{-5}$ | $1.15\times10^{-3}$ |
| 2.0 | global-baseline | 0.730 ± 1.740 | 76 | [103.704, 1054.630] | 1.977 | 0.023 | $5.94\times10^{-4}$ | $1.86\times10^{-2}$ |
| | plateau-window | 2.001 ± 0.237 | 25 | [150.301, 315.719] | 1.986 | 0.014 | $6.78\times10^{-5}$ | $2.93\times10^{-3}$ |
| 2.5 | global-baseline | 0.576 ± 2.097 | 76 | [97.484, 991.376] | 2.464 | 0.036 | $1.61\times10^{-3}$ | $4.36\times10^{-2}$ |
| | plateau-window | 2.502 ± 0.236 | 25 | [155.022, 325.635] | 2.487 | 0.013 | $1.05\times10^{-4}$ | $4.86\times10^{-3}$ |
| 3.0 | global-baseline | -0.021 ± 2.236 | 76 | [88.846, 903.538] | 2.933 | 0.067 | $3.60\times10^{-3}$ | $9.23\times10^{-2}$ |
| | plateau-window | 3.005 ± 0.278 | 25 | [150.301, 315.719] | 2.986 | 0.014 | $2.14\times10^{-4}$ | $1.16\times10^{-2}$ |
| 4.0 | global-baseline | -0.412 ± 2.346 | 76 | [0.010, 0.102] | -1.721 | 5.721 | $6.56\times10^{-3}$ | $1.00\times10^{-1}$ |
| | plateau-window | 3.952 ± 0.275 | 25 | [155.022, 325.635] | 3.928 | 0.072 | $3.36\times10^{-4}$ | $2.30\times10^{-2}$ |

The plateau-window protocol should therefore be understood as a finite-size regularization of the long-time fitting problem. It does not change the theoretical relation between the decay slope of $S(\infty) - S(t)$ and the scaling exponent. Instead, it provides a reproducible way to select the time interval over which that relation is applied. In the present framework, the spreadability method provides the dynamic estimate $\alpha_t$, which serves as one of the two primary numerical inputs to the joint empirical estimator.

### D. Joint empirical estimates and finite-size stability

After analyzing the three method-specific outputs, we combine the available estimates using the joint empirical estimator defined in Sec. IV D. Table VII summarizes the results for the nine $N_p = 200$ target systems. The number-variance method gives Class-like diagnoses consistent with the prescribed target regimes. For $\alpha_{\text{theory}} = 0.3$, 0.5 and 0.7, the systems are identified as Class III-like, so $\alpha_{\text{NV}}$ is included as an additional numerical contribution. For $\alpha_{\text{theory}} = 1.0$, the number-variance method gives a Class II-like diagnosis. For $\alpha_{\text{theory}} =$ 1.5, 2.0, 2.5, 3.0 and 4.0, it gives a Class I-like diagnosis. In these cases, number-variance contributes diagnostic information but does not enter the numerical average.

**TABLE VII.** Joint empirical estimates for the nine $N_p = 200$ target systems. The column $\sigma_N^2(R)$ class gives the Class-like diagnosis from the number-variance method. The quantities $\alpha_k$, $\alpha_t$, and $\alpha_{\text{NV}}$ are the estimates obtained from the structure-factor, spreadability, and number-variance methods, respectively. The joint estimate is $\alpha_{\text{joint}}$, and $u_{\text{joint}}$ measures the internal dispersion of the participating method estimates. Here $|\Delta\alpha_{\text{joint}}| = |\alpha_{\text{joint}} - \alpha_{\text{theory}}|$, and $\epsilon_{\text{joint}} = |\Delta\alpha_{\text{joint}}| / \alpha_{\text{theory}} \times 100\%$. A dash indicates that number-variance contributes only diagnostic information.

| $\alpha_{\text{theory}}$ | $\sigma_N^2(R)$ class | $\alpha_k$ | $\alpha_t$ | $\alpha_{NV}$ | $\alpha_{\text{joint}}$ | $u_{\text{joint}}$ | $\lvert\Delta\alpha_{\text{joint}}\rvert$ | $\epsilon_{\text{joint}}$ |
|---|---|---|---|---|---|---|---|---|
| 0.3 | Class III-like | 0.316 | 0.314 | 0.309 | 0.313 | 0.003 | 0.013 | 4.33% |
| 0.5 | Class III-like | 0.525 | 0.507 | 0.492 | 0.508 | 0.013 | 0.008 | 1.60% |
| 0.7 | Class III-like | 0.735 | 0.705 | 0.658 | 0.699 | 0.032 | 0.001 | 0.10% |
| 1.0 | Class II-like | 1.047 | 1.039 | - | 1.043 | 0.004 | 0.043 | 4.30% |
| 1.5 | Class I-like | 1.550 | 1.471 | - | 1.511 | 0.039 | 0.011 | 0.70% |
| 2.0 | Class I-like | 2.089 | 1.986 | - | 2.038 | 0.052 | 0.038 | 1.88% |
| 2.5 | Class I-like | 2.680 | 2.487 | - | 2.584 | 0.097 | 0.084 | 3.34% |
| 3.0 | Class I-like | 3.119 | 2.986 | - | 3.053 | 0.067 | 0.053 | 1.75% |
| 4.0 | Class I-like | 4.205 | 3.928 | - | 4.067 | 0.139 | 0.067 | 1.66% |

The joint estimates remain close to the prescribed target exponents across all nine systems. In the Class III-like cases, $\alpha_{\text{joint}}$ combines the reciprocal-space estimate $\alpha_k$, the dynamic estimate $\alpha_t$, and the real-space estimate $\alpha_{\text{NV}}$. In the Class II-like and Class I-like cases, $\alpha_{\text{joint}}$ is obtained from $\alpha_k$ and $\alpha_t$ only. This conditional rule follows the capability boundary of the number-variance method and avoids forcing a numerical number-variance inversion in regimes where the leading large-$R$ behavior no longer distinguishes different values of $\alpha > 1$.

Table VII also shows that $u_{\text{joint}}$ remains moderate for all target systems. This quantity should be interpreted as internal consistency rather than as a statistical uncertainty. For small $\alpha_{\text{theory}}$, the participating method estimates are close to one another, giving small $u_{\text{joint}}$. For larger $\alpha_{\text{theory}}$, the difference between the $S(k)$ and spreadability methods becomes more visible, and $u_{\text{joint}}$ increases. This behavior is expected because finite-size and preasymptotic effects affect the two methods differently. The joint estimate is therefore useful not because it removes method-level differences, but because it summarizes them under a transparent and reproducible rule.

Although Table VII uses an equal-weight empirical estimator, the same framework can be extended to a calibrated weighted form. For the Class I-like and Class II-like cases, where the number-variance method does not provide a numerical exponent, one may write $\alpha_{\text{joint}}(w) = w\alpha_k + (1-w)\alpha_t$. Using the six Class I-like and Class II-like benchmark entries in Table VII to fit a single optimal weight gives $w \simeq 0.19$, corresponding to a larger weight on the spreadability estimate. This calibrated weighted form reduces the mean absolute deviation of these six cases from 0.049 to 0.019, the root-mean-square deviation from 0.054 to 0.022, and the maximum absolute deviation from 0.084 to 0.041. If the three Class III-like cases are kept in the equal-weight three-method form, the overall mean absolute deviation over all nine benchmark systems decreases from 0.035 to 0.015. However, this improvement uses the known benchmark values of $\alpha_{\text{theory}}$ to calibrate the weight. Since $\alpha_{\text{theory}}$ is generally unavailable in applications to unknown systems, we use the equal-weight estimator as the default choice. This avoids target-dependent tuning and keeps $u_{\text{joint}}$ as a transparent measure of method-level dispersion.

**TABLE VIII.** Joint empirical estimates for the target system with $\alpha_{\text{theory}} = 3.0$ at different $N_p$. All systems are diagnosed as Class I-like by the number-variance method, so $\alpha_{\text{NV}}$ does not contribute to the numerical average. The quantities $\alpha_k$ and $\alpha_t$ are the structure-factor and spreadability estimates, respectively. The joint estimate is $\alpha_{\text{joint}}$, and $u_{\text{joint}}$ measures the internal dispersion of the participating method estimates. Here $|\Delta\alpha_{\text{joint}}| = |\alpha_{\text{joint}} - \alpha_{\text{theory}}|$, and $\epsilon_{\text{joint}} = |\Delta\alpha_{\text{joint}}|/\alpha_{\text{theory}} \times 100\%$.

| $N_p$ | $\sigma_N^2(R)$ class | $\alpha_k$ | $\alpha_t$ | $\alpha_{NV}$ | $\alpha_{\text{joint}}$ | $u_{\text{joint}}$ | $\lvert\Delta\alpha_{\text{joint}}\rvert$ | $\epsilon_{\text{joint}}$ |
|---|---|---|---|---|---|---|---|---|
| 100 | Class I-like | 3.031 | 2.978 | - | 3.005 | 0.027 | 0.005 | 0.15% |
| 200 | Class I-like | 3.119 | 2.986 | - | 3.053 | 0.067 | 0.053 | 1.75% |
| 300 | Class I-like | 3.174 | 2.957 | - | 3.066 | 0.109 | 0.066 | 2.18% |
| 500 | Class I-like | 3.056 | 2.924 | - | 2.990 | 0.066 | 0.010 | 0.33% |
| 800 | Class I-like | 3.053 | 2.934 | - | 2.994 | 0.060 | 0.007 | 0.22% |
| 1000 | Class I-like | 3.124 | 2.970 | - | 3.047 | 0.077 | 0.047 | 1.57% |

We further examine finite-size stability by varying $N_p$ for the target system with $\alpha_{\text{theory}} = 3.0$. The results are summarized in Table VIII. All systems in this test are diagnosed as Class I-like by the number-variance method, so $\alpha_{\text{NV}}$ does not enter the numerical average. The joint estimate is therefore obtained from $\alpha_k$ and $\alpha_t$.

The joint estimates remain close to the prescribed target exponent over the tested finite-size range. Specifically, $\alpha_{\text{joint}}$ varies from 2.990 to 3.066, with a maximum absolute deviation of 0.066. The purpose of this test is not to extrapolate to the thermodynamic limit. Rather, it shows that, within the small-to-intermediate finite-size range considered here, the method-specific estimates can be organized into a stable finite-size effective estimate.

Taken together, Tables VII and VIII support the main interpretation of the proposed framework. The $S(k)$ and spreadability methods provide the two primary numerical estimates. The number-variance method provides a capability-aware Class-like diagnosis and contributes a numerical reference only in Class III-like cases. The resulting $\alpha_{\text{joint}}$ is therefore a finite-size empirical summary of the available method outputs, not a strict reconstruction of the thermodynamic-limit asymptotic exponent. The internal consistency $u_{\text{joint}}$ provides additional information about method-level agreement and helps identify cases where the effective exponent is more sensitive to finite-size method differences.

### E. External validation using Gaussian pair statistics

To test whether the proposed finite-size protocol depends on the prescribed-$S(k)$ construction used in the main benchmark, we further considered an independently generated hyperuniform system with Gaussian pair statistics. This additional test examines whether the same analysis workflow remains effective for a system whose small-$k$ behavior is not imposed through direct target-$S(k)$ optimization. Specifically, fixed-$N$, fixed-density, periodic configurations were generated following the finite-periodic sampling procedure described in Ref. [58], with $N_p = 200$, $N_c = 100$, and $\rho = 1.0$. This system corresponds to the two-dimensional one-component plasma at $\Gamma = 2$. In the thermodynamic-limit bulk regime, it has the total correlation function

$$h(r) = g_2(r) - 1 = -\exp(-\pi\rho r^2) \tag{32}$$

and the corresponding structure factor is

$$S(k) = 1 + \rho\tilde{h}(k) = 1 - \exp\left[-\frac{k^2}{4\pi\rho}\right] \tag{33}$$

The small-wavenumber expansion is therefore

$$S(k) = \frac{k^2}{4\pi\rho} - \frac{k^4}{2(4\pi\rho)^2} + O(k^6) \tag{34}$$

Thus, the asymptotic hyperuniformity exponent is $\alpha_{\text{theory}} = 2$. This system provides a useful external validation case for two reasons. First, it is not generated by directly optimizing a prescribed target structure factor. Second, its structure factor is not a pure power law over the finite low-$k$ range, but contains an explicit negative $k^4$ correction. It therefore tests whether the proposed finite-size protocol can still extract a stable effective exponent when the accessible data contain finite-window curvature.

We applied the same protocol to $N_p = 200$ and $N_c = 100$ configurations at $\rho = 1.0$. The results are summarized in Table IX. The regularized structure-factor method gives $\alpha_k = 2.042$, which is close to the theoretical value. The number-variance method gives $\bar{p}_{\text{eff}} = 0.980$, corresponding to a Class I-like diagnosis. This is consistent with the expected behavior of a two-dimensional hyperuniform system with $\alpha > 1$, for which the leading large-$R$ number-variance growth is near-linear. Therefore, the number-variance output is used only as a Class-like diagnosis and is not included as a numerical exponent estimate.

**TABLE IX.** External validation on the periodic OCP/Gaussian pair-statistics system. The system corresponds to a fixed-$N$, fixed-density, periodic two-dimensional one-component plasma at $\Gamma = 2$, whose thermodynamic-limit bulk structure factor is $S(k) = 1 - \exp[-k^2/(4\pi\rho)]$. The number-variance method gives a Class I-like diagnosis and is not included in the numerical joint average.

| $\alpha_{\text{theory}}$ | $\bar{p}_{\text{eff}}$ | $\sigma_N^2(R)$ class | $\alpha_k$ | $\alpha_t$ | $\alpha_{\text{joint}}$ | $u_{\text{joint}}$ | $\lvert\Delta\alpha_{\text{joint}}\rvert$ | $\epsilon_{\text{joint}}$ |
|---|---|---|---|---|---|---|---|---|
| 2.0 | 0.980 | Class I-like | 2.042 | 1.888 | 1.965 | 0.077 | 0.035 | 1.75% |

The spreadability method gives $\alpha_t = 1.888$, slightly below the theoretical value. This finite-size deviation is consistent with the curvature of the Gaussian structure factor over the accessible low-$k$ window. Combining the available numerical estimates from $S(k)$ and spreadability gives $\alpha_{\text{joint}} = (\alpha_k + \alpha_t)/2 = 1.965$, with $u_{\text{joint}} = 0.077$, $\lvert\Delta\alpha_{\text{joint}}\rvert = 0.035$, and $\epsilon_{\text{joint}} = 1.75\%$. These results show that the proposed protocol remains accurate and internally consistent for an independently generated hyperuniform system beyond the target-$S(k)$ benchmark.

# VI. CONCLUSIONS

In this work, we developed a unified protocol for extracting effective scaling exponents in finite-size hyperuniform systems. The motivation is that the scaling exponent $\alpha$ is a useful descriptor of how long-wavelength density fluctuations are suppressed, but its extraction from finite samples is strongly affected by the accessible fitting window, finite-size contamination, and the different capabilities of the available structural methods. We therefore formulated the problem as a finite-window and method-consistency problem, rather than as a direct reconstruction of the thermodynamic-limit asymptotic exponent.

The proposed framework combines three complementary methods. In the structure-factor method, regularized low-$k$ window screening selects contiguous fitting windows by balancing log-log fitting quality and boundary perturbation stability. This reduces the dependence on a fixed empirical cutoff and provides a reproducible low-$k$ effective estimate. In the number-variance method, tail-platform identification is used for Class-like diagnosis. This method contributes numerical exponent information only in Class III-like cases, where the large-$R$ behavior retains exponent information through $\sigma_N^2(R) \sim R^{2-\alpha}$ [1, 3-4, 46]. In the spreadability method, the plateau-window protocol selects a shared long-time fitting window for the spreadability excess $S(\infty) - S(t)$. This regularizes the dynamic fitting problem without changing the theoretical relation between the long-time decay and the scaling exponent [15-17].

We tested the protocol using two-dimensional target hyperuniform point configurations generated under prescribed $S(k)$ constraints [54]. This benchmark allowed the target exponent, particle number, number of configurations, and number density to be controlled. The results show that the regularized $S(k)$ method reduces the sensitivity to empirical low-$k$ window selection, while still retaining finite-size deviations for larger $\alpha_{\text{theory}}$. The number-variance method provides stable Class-like diagnosis and contributes numerical estimates only in the Class III-like regime. The spreadability method, when analyzed through the plateau-window protocol, gives much more stable configuration-level estimates than an open global-window search, especially for larger target exponents.

The method-specific estimates were summarized through a joint empirical estimator. The $S(k)$ and spreadability methods serve as the primary numerical inputs, while the number-variance method enters the numerical average only when its Class-like diagnosis indicates that exponent information is available. The internal consistency $u_{\text{joint}}$ quantifies the dispersion among the participating method estimates. It is not a statistical confidence interval or an asymptotic uncertainty. In the benchmark tests, the joint estimates remained close to the prescribed target exponents across the tested $\alpha_{\text{theory}}$ values and remained stable over the finite-size range examined for $\alpha_{\text{theory}} = 3.0$.

The present benchmark is a controlled finite-size methodology test based on prescribed-$S(k)$ configurations, while the additional Gaussian pair-statistics validation indicates that the protocol also remains accurate for an independently generated hyperuniform system not constructed by directly optimizing a prescribed target $S(k)$.

Future applications may also adapt the method-specific window selection criteria to anisotropic systems, multicomponent point patterns, and two-phase media with more complex inclusion shapes. Such extensions would help connect finite-size hyperuniformity diagnostics with practical structural characterization.

## ACKNOWLEDGMENTS

This work has been supported by the National Natural Science Foundation of China (Grants No. 11274200, 12274255) , the Natural Science Foundation of Shandong Province, China (Grant No. ZR2022MA055), and City University of Hong Kong (Grant No. 9610532).

## SUPPLEMENTARY MATERIAL

The supplementary material contains two-phase spectral-density validation for the disk-decoration step and sensitivity tests for the control parameters used in the finite-size effective-exponent extraction protocol. These tests examine the effect of the decoration fraction, the regularized low-$k$ window screening in the structure-factor method, the tail-platform identification in the number-variance method, and the plateau-window protocol for the spreadability excess $S(\infty) - S(t)$.

## AUTHOR DECLARATIONS

### Conflict of Interest

The authors have no conflicts to disclose.

### Data Availability

The source code and a representative example dataset associated with this study are available in a GitHub repository at https://github.com/JianxiangTian/finite-size-hyperuniform-alpha. The repository contains the finite-size effective-exponent extraction workflow, method-specific analysis scripts, parameter files, documentation, and an example dataset for demonstrating the complete analysis pipeline from point configurations to method-specific and joint empirical estimates. Additional benchmark configurations used in the paper are available from the corresponding author upon reasonable request.

## APPENDIX: Finite-size interpretation of the spreadability fitting window

In a two-dimensional isotropic system, the spreadability excess can be used to characterize the large-scale structural response in a long-time diffusion process. We denote $E(\tau) = S(\infty) - S(t)$, where $\tau = Dt/R_d^2$, $t = R_d^2\tau/D$. Here, $D$ is the diffusion coefficient, and $R_d$ is the disk radius used to decorate the point configuration into a two-phase medium. Ignoring normalization constants that do not affect the scaling exponent, $E(\tau)$ can be expressed through a Fourier-space integral of the corresponding spectral density $\tilde{\chi}(k)$ as [15-17]

$$E(\tau) \simeq \frac{1}{2\pi}\int_0^\infty \tilde{\chi}(k)\exp(-R_d^2\tau k^2)k\,dk \tag{A1}$$

Here, $\tilde{\chi}(k)$ denotes the spectral density of the decorated two-phase medium used in the diffusion calculation. In the present protocol, the point configuration is decorated by identical disks and phase 2 is taken as the union of the disk domains [4, 17]. We use the spreadability method as a dynamic finite-size probe of the large-scale spectral behavior of this decorated medium. For compact notation, $S(k)$ below denotes the low-$k$ spectral quantity that controls the long-time decay of $E(\tau)$.

If the small-wavenumber region satisfies

$$S(k) \simeq Ak^\alpha, \qquad 0 < k < k_0 \tag{A2}$$

then, in the long-time regime dominated by small wavenumbers, Eq. (A1) can be approximated as

$$E(\tau) \simeq \frac{A}{2\pi}\int_0^\infty k^{\alpha+1}\exp(-R_d^2\tau k^2)\,dk \tag{A3}$$

Writing the upper limit as infinity does not imply that the actual $S(k)$ follows $Ak^\alpha$ over the entire wavenumber range. The physical reason is that, in the long-time regime, the Gaussian kernel $\exp(-R_d^2\tau k^2)$ strongly suppresses large-$k$ contributions. The dominant contribution to the integral then comes from the small-wavenumber region.

Using

$$\int_0^\infty k^{\alpha+1}\exp(-R_d^2\tau k^2)\,dk = \frac{1}{2}(R_d^2\tau)^{-(1+\frac{\alpha}{2})}\Gamma\left(1+\frac{\alpha}{2}\right) \tag{A4}$$

we obtain

$$E(\tau) \simeq \frac{A}{4\pi}R_d^{-(2+\alpha)}\Gamma\left(1+\frac{\alpha}{2}\right)\tau^{-1-\frac{\alpha}{2}} \tag{A5}$$

Thus, in two dimensions,

$$S(\infty) - S(t) \sim \tau^{-1-\frac{\alpha}{2}} \tag{A6}$$

If, over a certain time interval,

$$E(\tau) \simeq C\tau^{-p} \tag{A7}$$

then

$$\log E(\tau) = \log C - p\log\tau \tag{A8}$$

If the slope of the log-log fit is denoted by $m$, then

$$m = -p \tag{A9}$$

In two dimensions,

$$p = 1 + \frac{\alpha}{2} \tag{A10}$$

Therefore,

$$m = -\left(1 + \frac{\alpha}{2}\right), \quad \alpha = -2m - 2 \tag{A11}$$

This shows that extracting $\alpha$ from an approximately linear segment in a log-log plot of $E(\tau)$ is not merely empirical. It follows from the correspondence between the small-wavenumber power law and the long-time diffusion decay. The difficulty is that, in a finite system, this approximately linear interval may not be wide, and it does not necessarily occur at the latest accessible times.

In the numerical examples considered here, the amplitude $A$ is obtained by fitting the actual finite-size $S_{\text{data}}(k)$ over a prescribed low-wavenumber power-law interval. The upper endpoint of this interval is denoted by $k_0$, and it satisfies $k_0 a = 2.5$ where $a$ is the ensemble-averaged nearest-neighbor distance. The quantities $A$ and $k_0$ are used only to construct theoretical reference curves and to interpret the effective time window. They are not used in the plateau-window selection protocol.

To clarify the finite-size effect on the above correspondence, we introduce two reference quantities. The first is the continuous ideal power-law prediction

$$E_{\text{theory}}(\tau) = \frac{A}{4\pi} R_d^{-(2+\alpha)} \Gamma\left(1 + \frac{\alpha}{2}\right) \tau^{-1-\frac{\alpha}{2}} \tag{A12}$$

The second reference quantity is the ideal power-law result evaluated on the same finite $k$-grid as the numerical data. Let $k_n$ be the finite wavenumber grid used for $S_{\text{data}}(k)$, and let $w_n$ denote the quadrature weights that include the radial measure $k\,dk$. Then

$$E_{\text{grid}}(\tau) \simeq \frac{1}{2\pi} \sum_n w_n A k_n^{\alpha} \exp(-R_d^2 \tau k_n^2) \tag{A13}$$

The role of $E_{\mathrm{grid}}(\tau)$ is to isolate the influence of the finite $k$-grid and the finite wavenumber range [48]. In practice, the range of $k_n$ is the same as that used for the actual data. Its lower bound is associated with the smallest nonzero wavenumber $k_{\mathrm{min}}$ in the finite periodic system, and its upper bound is the largest available wavenumber used in the numerical integration. In contrast, $E_{\mathrm{data}}(\tau)$ is obtained by numerically integrating the actual finite-size $S_{\mathrm{data}}(k)$. Therefore, deviations between $E_{\mathrm{grid}}(\tau)$ and $E_{\mathrm{theory}}(\tau)$ indicate that the finite $k$-grid itself already affects the integral. Deviations between $E_{\mathrm{data}}(\tau)$ and $E_{\mathrm{grid}}(\tau)$ indicate that the actual $S_{\mathrm{data}}(k)$ does not fully follow $Ak^{\alpha}$ over the effective wavenumber range.

The Gaussian kernel in Eq. (A1) determines the effective wavenumber scale associated with the dimensionless time $\tau$. Roughly speaking, the dominant contribution comes from the region $R_d^2 \tau k^2 \lesssim 1$. Thus, at the order-of-magnitude level, we may define

$$k_{\mathrm{eff}} \sim \frac{1}{R_d \sqrt{\tau}} \tag{A14}$$

As $\tau$ increases, the integral is progressively concentrated at smaller wavenumbers. If the actual $S_{\mathrm{data}}(k)$ follows $Ak^{\alpha}$ only in the interval $0 < k < k_0$, then Eq. (A5) requires $k_{\mathrm{eff}} \ll k_0$. This gives the lower time condition

$$\tau \gg \frac{1}{R_d^2 k_0^2} \tag{A15}$$

On the other hand, in a finite periodic system, the smallest nonzero wavenumber is approximately

$$k_{\mathrm{min}} = 2\pi / L \tag{A16}$$

The finite system does not contain structural information for $0 < k < k_{\mathrm{min}}$. If $\tau$ is so large that $k_{\mathrm{eff}}$ becomes comparable to or smaller than $k_{\mathrm{min}}$, the integral requires small-wavenumber information that the finite system cannot provide. The power-law relation is then affected by finite-size contamination. Therefore, we also require $k_{\mathrm{min}} \ll k_{\mathrm{eff}}$, which gives the upper time condition

$$\tau \ll \frac{1}{R_d^2 k_{\mathrm{min}}^2} \tag{A17}$$

Combining the two conditions gives the finite-size admissible range

$$\frac{1}{R_d^2 k_0^2} \ll \tau \ll \frac{1}{R_d^2 k_{\mathrm{min}}^2} \tag{A18}$$

Equation (A18) shows that the fitting interval for $S(\infty) - S(t)$ should be neither too early nor too late. At early times, the integral is still influenced by larger-wavenumber structures, where the actual $S_{\mathrm{data}}(k)$ usually deviates from $Ak^{\alpha}$. At very late times, the integral enters a finite-size-controlled regime, where structural information for $0 < k < k_{\min}$ is absent. The curve may then deviate from the ideal power law or bend downward rapidly. Therefore, the effective power-law fitting interval is expected to appear in an intermediate time range. Equation (A18) should be interpreted as an order-of-magnitude condition rather than a sharp mathematical boundary. The actual stable window also depends on the spectral shape, the finite $k$-grid, and the width of the low-$k$ power-law region.

To quantify the agreement between the actual integral and the ideal power-law prediction, we define

$$R_{\mathrm{data}}(\tau) = \frac{E_{\mathrm{data}}(\tau)}{E_{\mathrm{theory}}(\tau)} \tag{A19}$$

and

$$R_{\mathrm{grid}}(\tau) = \frac{E_{\mathrm{grid}}(\tau)}{E_{\mathrm{theory}}(\tau)} \tag{A20}$$

When $R_{\mathrm{data}}(\tau)$ is close to unity, the actual finite-size data agree with the continuous power-law prediction over that time interval. When $R_{\mathrm{grid}}(\tau)$ is close to unity, the finite $k$-grid itself can still reproduce the continuous power-law prediction. Comparing the two ratios helps separate the influence of the finite $k$-grid from the influence of deviations of the actual $S_{\mathrm{data}}(k)$ from the ideal power law.

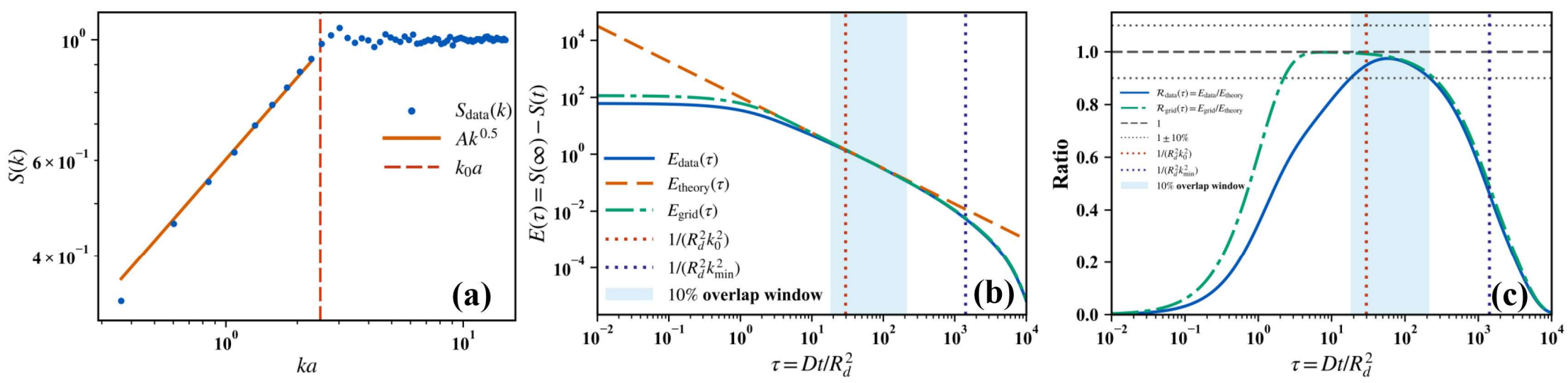

**FIG. 6.** Local power-law approximation of $S_{\mathrm{data}}(k)$ and its effect on the spreadability excess for a finite-size sample with $\alpha = 0.5$. The example uses $N = 200$ and $\alpha = 0.5$. Panel (a) compares the actual $S_{\mathrm{data}}(k)$ with the fitted form $Ak^{0.5}$. The horizontal axis is $ka$, and the vertical dashed line marks the right endpoint of the prescribed local power-law reference interval, $k_0 a = 2.5$. Panel (b) compares $E_{\mathrm{data}}(\tau)$, $E_{\mathrm{theory}}(\tau)$, and $E_{\mathrm{grid}}(\tau)$. Here, $E_{\mathrm{data}}(\tau)$ is obtained by numerically integrating the actual finite-size $S_{\mathrm{data}}(k)$, $E_{\mathrm{theory}}(\tau)$ is the continuous power-law prediction, and $E_{\mathrm{grid}}(\tau)$ is obtained by summing the ideal power law over the same finite $k$-grid. The two vertical lines mark $1/(R_d^2 k_0^2)$ and $1/(R_d^2 k_{\min}^2)$, respectively. The blue shaded region indicates the 10% overlap window. Panel (c) shows the ratios $R_{\mathrm{data}}(\tau) = E_{\mathrm{data}}(\tau)/E_{\mathrm{theory}}(\tau)$ and $R_{\mathrm{grid}}(\tau) = E_{\mathrm{grid}}(\tau)/E_{\mathrm{theory}}(\tau)$. The horizontal dashed lines mark 1 and 1±10%. The figure shows that the actual data approach the ideal power-law prediction only within a finite intermediate time window, while both early and late times show deviations.

Figure 6 shows the finite-size sample with $\alpha = 0.5$. In Figure 6(a), the actual $S_{\mathrm{data}}(k)$ follows a local power-law approximation in the interval $ka \leq 2.5$ and then gradually enters a plateau at larger wavenumbers. This confirms that Eq. (A2) is not a relation over the full wavenumber range, but only a local approximation in the low-$k$ region. Figure 6(b) shows that $E_{\mathrm{data}}(\tau)$ is close to $E_{\mathrm{theory}}(\tau)$ only in an intermediate time interval. The ratio curves in Figure 6(c) further show that $R_{\mathrm{data}}(\tau)$ is close to unity within the 10% overlap window highlighted by the blue shading, while both early and late times deviate from unity. At the same time, $R_{\mathrm{grid}}(\tau)$ remains close to unity over a wider time range, indicating that the finite $k$-grid itself is not the only reason for the narrowing of the actual overlap window.

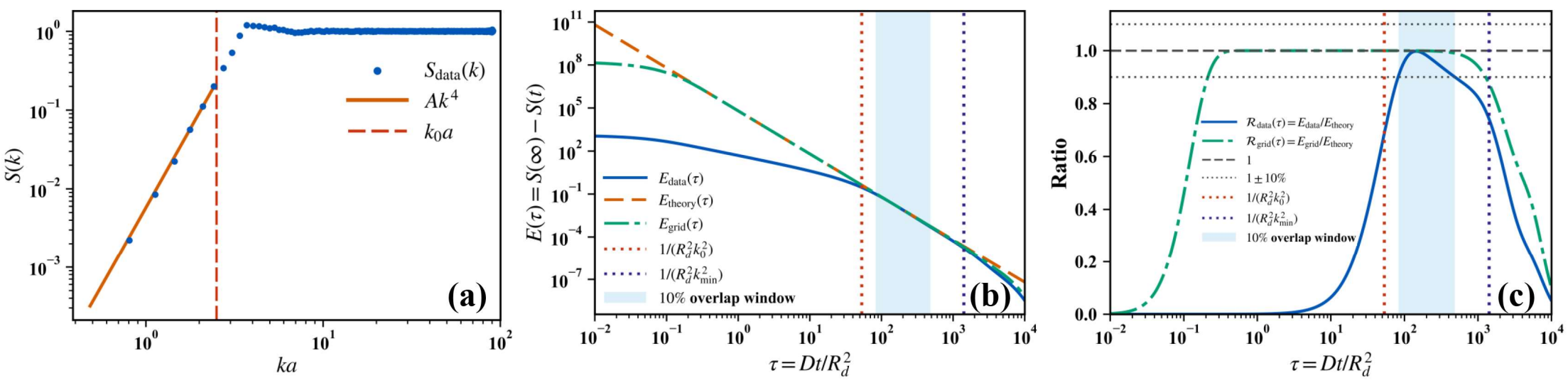

**FIG. 7.** Window behavior for a larger scaling exponent in a finite-size sample with $\alpha = 4.0$. The example uses $N = 200$ and $\alpha = 4.0$. Panel (a) compares the actual $S_{\mathrm{data}}(k)$ with the fitted form $Ak^4$. The horizontal axis is $ka$, and the vertical dashed line marks the right endpoint of the prescribed local power-law reference interval, $k_0 a = 2.5$. Panel (b) compares $E_{\mathrm{data}}(\tau)$, $E_{\mathrm{theory}}(\tau)$, and $E_{\mathrm{grid}}(\tau)$. The blue shaded region indicates the 10% overlap window. Panel (c) shows the ratios $R_{\mathrm{data}}(\tau) = E_{\mathrm{data}}(\tau)/E_{\mathrm{theory}}(\tau)$ and $R_{\mathrm{grid}}(\tau) = E_{\mathrm{grid}}(\tau)/E_{\mathrm{theory}}(\tau)$. Compared with the $\alpha = 0.5$ case, the actual overlap window for $\alpha = 4.0$ is narrower overall, indicating that, under the same finite-size sample and diagnostic criterion used here, larger $\alpha$ is more strongly constrained by finite size and window boundaries.

Figure 7 shows the finite-size sample with $\alpha = 4.0$. As in the $\alpha = 0.5$ case, the actual $S_{\mathrm{data}}(k)$ in Figure 7(a) approaches $Ak^4$ only over a finite low-wavenumber interval and then enters a plateau. The difference is that Figure 7(b) and 7(c) show that the overlap window between $E_{\mathrm{data}}(\tau)$ and $E_{\mathrm{theory}}(\tau)$ occurs later and is limited to a narrower finite time range. In particular, under the 10% tolerance criterion, $R_{\mathrm{data}}(\tau)$ is close to unity only over an intermediate time window. This indicates that the effective power-law interval is more strongly constrained by finite-size and window-boundary effects for larger $\alpha$.

This behavior can be understood from the decay exponent in two dimensions. From Eq. (A10), $p = 1 + \alpha/2$. For $\alpha = 0.5$, $E(\tau) \sim \tau^{-1.25}$, whereas for $\alpha = 4.0$, $E(\tau) \sim \tau^{-3}$. Thus, under the same finite-size sample and the same diagnostic criterion, a larger $\alpha$ leads to a faster decay of $E(\tau)$, making the power-law signal more likely to approach the finite-size tail or the numerical-error regime. By contrast, a smaller $\alpha$ corresponds to a slower long-time decay and often retains a wider effective power-law window in finite systems. This statement should be interpreted as an empirical observation for the finite-size samples studied here, not as a universal theorem for all systems.

We define the overlap interval between the actual integral and the ideal power-law prediction by

$$\left|\frac{E_{\text{data}}(\tau)}{E_{\text{theory}}(\tau)} - 1\right| \le \epsilon \tag{A21}$$

where $\epsilon$ is the tolerance. Table X reports the overlap intervals for $\epsilon = 5\%$, $10\%$, and $20\%$. The results show that the actual overlap window for $\alpha = 0.5$ is generally wider than that for $\alpha = 4.0$. For example, at the $10\%$ tolerance level, the actual overlap interval is [18.158, 215.237] for $\alpha = 0.5$, with a span of $1.074$ decades, whereas it is [83.157, 485.129] for $\alpha = 4.0$, with a span of $0.766$ decades. This indicates that the effective power-law window is more easily compressed by finite-size and window-boundary effects for the larger exponent. This difference should be understood as an empirical behavior under the finite sample size and diagnostic standard used here, not as a strict universal conclusion.

**TABLE X.** Overlap intervals between the actual finite-size integral and the ideal power-law prediction. Here, $I_{\text{data}}$ denotes the longest continuous interval in which $R_{\text{data}}(\tau) = E_{\text{data}}(\tau)/E_{\text{theory}}(\tau)$ lies within $1 \pm \epsilon$, and $I_{\text{grid}}$ denotes the longest continuous interval in which $R_{\text{grid}}(\tau) = E_{\text{grid}}(\tau)/E_{\text{theory}}(\tau)$ lies within $1 \pm \epsilon$. The quantities $\Delta_{\text{data}}$ and $\Delta_{\text{grid}}$ are the spans of $I_{\text{data}}$ and $I_{\text{grid}}$ on the logarithmic dimensionless-time scale, defined as $\log_{10}(\tau_{\text{end}}/\tau_{\text{start}})$.

| $\alpha$ | $\varepsilon$ | $I_{\text{data}}$ | $\Delta_{\text{data}}$ | $I_{\text{grid}}$ | $\Delta_{\text{grid}}$ |
|---|---|---|---|---|---|
| 0.5 | 5% | [30.504, 119.563] | 0.593 | [2.855, 134.947] | 1.675 |
| 0.5 | 10% | [18.158, 215.237] | 1.074 | [2.241, 247.167] | 2.043 |
| 0.5 | 20% | [8.937, 422.457] | 1.675 | [1.613, 468.639] | 2.463 |
| 4.0 | 5% | [97.159, 278.970] | 0.458 | [0.254, 919.824] | 3.559 |
| 4.0 | 10% | [83.157, 485.129] | 0.766 | [0.213, 1255.664] | 3.770 |
| 4.0 | 20% | [66.417, 1131.924] | 1.232 | [0.173, 1805.389] | 4.018 |

Table X also shows that, for the same tolerance, the overlap interval for the finite-grid ideal result is generally wider than that for the actual integral. For example, for $\alpha = 4.0$, the $10\%$ tolerance gives a span of $3.770$ decades for $I_{\text{grid}}$, but only $0.766$ decades for $I_{\mathrm{data}}$. For $\alpha = 0.5$, the $10\%$ tolerance gives a span of $2.043$ decades for $I_{\mathrm{grid}}$, but $1.074$ decades for $I_{\mathrm{data}}$. This means that the instability of the actual data is not caused only by the minimum-wavenumber cutoff or by the finite $k$-grid itself. It also comes from the finite width of the power-law region in the actual $S_{\mathrm{data}}(k)$.

The plateau-window method proposed in this work is based on this interpretation. It does not modify the theoretical relation in Eq. (A6). Instead, it converts the finite-size window-selection problem into a stability-identification problem. By using a diffusion-length constraint, local effective-exponent plateau identification, and a shared-window fit, the method restricts the fitting interval to a physically interpretable intermediate time range. This reduces the risk that an open search is attracted to an early preasymptotic regime or to a late finite-size tail. Therefore, the extracted $\alpha$ should be understood as a stably identifiable effective scaling exponent in a finite-size system, rather than a direct reconstruction of the strict asymptotic exponent in an infinite system.

# REFERENCES

[1] S. Torquato, Hyperuniform states of matter, Phys. Rep. **745**, 1-95 (2018).

[2] H. W. Haslach, Jr., Random Heterogeneous Materials: Microstructure and Macroscopic Properties, Appl. Mech. Rev. **55**, B62-B63 (2002).

[3] S. Torquato and F. H. Stillinger, Local density fluctuations, hyperuniformity, and order metrics, Phys. Rev. E **68**, 041113 (2003).

[4] C. E. Zachary and S. Torquato, Hyperuniformity in point patterns and two-phase random heterogeneous media, J. Stat. Mech.:Theory Exp. **2009**, P12015 (2009).

[5] G. J. Aubry, L. S. Froufe-Pérez, U. Kuhl, O. Legrand, F. Scheffold and F. Mortessagne, Experimental Tuning of Transport Regimes in Hyperuniform Disordered Photonic Materials, Phys. Rev. Lett. **125**, 127402 (2020).

[6] M. Florescu, S. Torquato and P. J. Steinhardt, Designer disordered materials with large, complete photonic band gaps, Proc. Natl. Acad. Sci. **106**, 20658-20663 (2009).

[7] L. S. Froufe-Pérez, M. Engel, J. J. Sáenz and F. Scheffold, Band gap formation and Anderson localization in disordered photonic materials with structural correlations, Proc. Natl. Acad. Sci. **114**, 9570-9574 (2017).

[8] M. A. Klatt, P. J. Steinhardt and S. Torquato, Wave propagation and band tails of two-dimensional disordered systems in the thermodynamic limit, Proc. Natl. Acad. Sci. **119**, e2213633119 (2022).

[9] W. Man, M. Florescu, K. Matsuyama, P. Yadak, G. Nahal, S. Hashemizad, E. Williamson, P. Steinhardt, S. Torquato and P. Chaikin, Photonic band gap in isotropic hyperuniform disordered solids with low dielectric contrast, Opt. Express **21**, 19972-19981 (2013).

[10] O. Christogeorgos, H. Zhang, Q. Cheng and Y. Hao, Extraordinary Directive Emission and Scanning from an Array of Radiation Sources with Hyperuniform Disorder, Phys. Rev. Appl. **15**, 014062 (2021).

[11] R. D. Batten, F. H. Stillinger and S. Torquato, Novel Low-Temperature Behavior in Classical Many-Particle Systems, Phys. Rev. Lett. **103**, 050602 (2009).

[12] R. D. Batten, F. H. Stillinger and S. Torquato, Interactions leading to disordered ground states and unusual low-temperature behavior, Phys. Rev. E **80**, 031105 (2009).

[13] Q. Le Thien, D. Mcdermott, C. J. O. Reichhardt and C. Reichhardt, Enhanced pinning for vortices in hyperuniform pinning arrays and emergent hyperuniform vortex configurations with quenched disorder, Phys. Rev. B **96**, 094516 (2017).

[14] D. Chen and S. Torquato, Designing disordered hyperuniform two-phase materials with novel physical properties, Acta Mater. **142**, 152-161 (2018).

[15] M. Skolnick and S. Torquato, Simulated diffusion spreadability for characterizing the structure and transport properties of two-phase materials, Acta Mater. **250**, 118857 (2023).

[16] S. Torquato, Diffusion spreadability as a probe of the microstructure of complex media across length scales, Phys. Rev. E **104**, 054102 (2021).

[17] H. Wang and S. Torquato, Dynamic Measure of Hyperuniformity and Nonhyperuniformity in Heterogeneous Media via the Diffusion Spreadability, Phys. Rev. Appl. **17**, 034022 (2022).

[18] E. Nocerino, Emergent properties and the multiscale characterization challenge in condensed matter, from crystals to complex materials: a review, J. Phys. D:Appl. Phys. **58**, 393001 (2025).

[19] J. S. Brauchart, P. J. Grabner and W. Kusner, Hyperuniform Point Sets on the Sphere: Deterministic Aspects, Constr. Approx. **50**, 45-61 (2019).

[20] J. S. Brauchart, P. J. Grabner, W. Kusner and J. Ziefle, Hyperuniform point sets on the sphere: probabilistic aspects, Monatsh. Math. **192**, 763-781 (2020).

[21] S. Ghosh and J. L. Lebowitz, Generalized Stealthy Hyperuniform Processes: Maximal Rigidity and the Bounded Holes Conjecture, Commun. Math. Phys. **363**, 97-110 (2018).

[22] S. Torquato, G. Zhang and M. De Courcy-Ireland, Hidden multiscale order in the primes, J. Phys. A:Math. Theor. **52**, 135002

(2019).

[23] I. Gudoshnikov, Y. Jiao, O. Makarenkov and D. Chen, Sweeping process approach to stress analysis in elastoplastic lattice spring models with applications to network materials, Phys. Rev. E **112**, 065501 (2025).

[24] Y. Jiao, Acoustic-Edge and Thermal Scaling in Disordered Hyperuniform Networks: A First Principles Theory, Phys. Rev. Lett. **136**, 137402 (2026).

[25] E. Newby, W. Shi, Y. Jiao, R. Albert and S. Torquato, Structural properties of hyperuniform Voronoi networks, Phys. Rev. E **111**, 034123 (2025).

[26] E. Newby, W. Shi, Y. Jiao, S. Torquato and R. Albert, From point patterns to networks: to what extent does the Delaunay triangulation reproduce key spatial and density information?, J. Complex Netw. **13**, cnaf040 (2025).

[27] J. Wang, Z. Sun, H. Chen, G. Wang, D. Chen, G. Chen, J. Shuai, M. Yang, Y. Jiao and L. Liu, Hyperuniform Networks of Active Magnetic Robotic Spinners, Phys. Rev. Lett. **134**, 248301 (2025).

[28] E. Ballestero, A. Duclos, A. Barbacci and V. Romero-García, Emergence of hyperuniformity from reaction-diffusion interactions in Turing patterns, Phys. Rev. E **112**, L062401 (2025).

[29] Z. Ge, The hidden order of Turing patterns in arid and semi-arid vegetation ecosystems, Proc. Natl. Acad. Sci. **120**, e2306514120 (2023).

[30] W. Hu, L. Cui, M. Delgado-Baquerizo, R. Solé, S. Kéfi, M. Berdugo, N. Xu, B. Wang, Q.-X. Liu and C. Xu, Causes and consequences of disordered hyperuniformity in global drylands, Proc. Natl. Acad. Sci. **122**, e2504496122 (2025).

[31] Y. Jiao, T. Lau, H. Hatzikirou, M. Meyer-Hermann, C. C. Joseph and S. Torquato, Avian photoreceptor patterns represent a disordered hyperuniform solution to a multiscale packing problem, Phys. Rev. E **89**, 022721 (2014).

[32] Z.-Q. Li, Q.-L. Lei and Y.-Q. Ma, Fluidization and anomalous density fluctuations in 2D Voronoi cell tissues with pulsating activity, Proc. Natl. Acad. Sci. **122**, e2421518122 (2025).

[33] S. Lin, Z. Du, X. Fang, G. Li and Y. M. Xie, Biological patterns and textures: From formation mechanisms to biomimetic applications, J. Mech. Behav. Biomed. Mater. **178**, 107392 (2026).

[34] Y. Liu, D. Chen, J. Tian, W. Xu and Y. Jiao, Universal Hyperuniform Organization in Looped Leaf Vein Networks, Phys. Rev. Lett. **133**, 028401 (2024).

[35] A. Mayer, V. Balasubramanian, T. Mora and A. M. Walczak, How a well-adapted immune system is organized, Proc. Natl. Acad. Sci. **112**, 5950-5955 (2015).

[36] N. Wang, Y. Tong, F. Liu, X. Li, Y. He and W. Fan, Turing instability-induced oscillations in coupled reaction-diffusion systems, Chin. Phys. B **34**, 038201 (2025).

[37] R. J. H. Ross, G. D. Masucci, C. Y. Lin, T. L. Iglesias, S. Reiter and S. Pigolotti, Hyperdisordered Cell Packing on a Growing Surface, Phys. Rev. X **15**, 021064 (2025).

[38] D. Chen, R. Samajdar, Y. Jiao and S. Torquato, Anomalous suppression of large-scale density fluctuations in classical and quantum spin liquids, Proc. Natl. Acad. Sci. **122**, e2416111122 (2025).

[39] Y. Lu, T. Zhu, Y. Guo, Y. Li and Z. Zheng, Orientation-Modulated Hyperuniformity in Frustrated Vicsek–Kuramoto Systems, Entropy **28**, 126 (2026).

[40] R. Maire and L. Chaix, Hyperuniformity and conservation laws in non-equilibrium systems, J. Chem. Phys. **163**, 214507 (2025).

[41] L. Zhong and S. Mao, Construction of digital realizations of anisotropic hyperuniform continuous random fields in two and three dimensions via generalized spectral filtering, Phys. Rev. E **112**, 044136 (2025).

[42] C. E. Maher and S. Torquato, Hyperuniformity scaling of maximally random jammed packings of two-dimensional binary disks, Phys. Rev. E **110**, 064605 (2024).

[43] J. Shang, Y. Wang, D. Pan, Y. Jin and J. Zhang, Jamming as a topological satisfiability transition with contact number hyperuniformity and criticality, Proc. Natl. Acad. Sci. **123**, e2517241123 (2026).

[44] C. E. Zachary, Y. Jiao and S. Torquato, Hyperuniform Long-Range Correlations are a Signature of Disordered Jammed Hard-

Particle Packings, Phys. Rev. Lett. **106**, 178001 (2011).

[45] A. Ikeda, L. Berthier and G. Parisi, Large-scale structure of randomly jammed spheres, Phys. Rev. E **95**, 052125 (2017).

[46] C. E. Zachary and S. Torquato, Anomalous local coordination, density fluctuations, and void statistics in disordered hyperuniform many-particle ground states, Phys. Rev. E **83**, 051133 (2011).

[47] A. Hitin-Bialus, C. E. Maher, P. J. Steinhardt and S. Torquato, Hyperuniformity classes of quasiperiodic tilings via diffusion spreadability, Phys. Rev. E **109**, 064108 (2024).

[48] Y. Liu, X. Li, J. Tian, X. Yan and G. Zhang, Impact of random spatial truncation and reciprocal-space binning on the detection of hyperuniformity in disordered systems, J. Chem. Phys. **164**, 094102 (2026).

[49] L. E. Silbert and M. Silbert, Long-wavelength structural anomalies in jammed systems, Phys. Rev. E **80**, 041304 (2009).

[50] O. U. Uche, F. H. Stillinger and S. Torquato, Constraints on collective density variables: Two dimensions, Phys. Rev. E **70**, 046122 (2004).

[51] S. Atkinson, G. Zhang, A. B. Hopkins and S. Torquato, Critical slowing down and hyperuniformity on approach to jamming, Phys. Rev. E **94**, 012902 (2016).

[52] D. Chen, E. Lomba and S. Torquato, Binary mixtures of charged colloids: a potential route to synthesize disordered hyperuniform materials, Phys. Chem. Chem. Phys. **20**, 17557-17562 (2018).

[53] S. Torquato, Structural characterization of many-particle systems on approach to hyperuniform states, Phys. Rev. E **103**, 052126 (2021).

[54] G. Zhang and S. Torquato, Realizable hyperuniform and nonhyperuniform particle configurations with targeted spectral functions via effective pair interactions, Phys. Rev. E **101**, 032124 (2020).

[55] Y. Jiao, Spectral leakage and masking effects in the measurement of hyperuniformity, J. Chem. Phys. **164**, 224106 (2026).

[56] R. Dreyfus, Y. Xu, T. Still, L. A. Hough, A. G. Yodh and S. Torquato, Diagnosing hyperuniformity in two-dimensional, disordered, jammed packings of soft spheres, Phys. Rev. E **91**, 012302 (2015).

[57] D. Hawat, G. Gautier, R. Bardenet and R. Lachièze-Rey, On estimating the structure factor of a point process, with applications to hyperuniformity, Stat. Comput. **33**, 61 (2023).

[58] H. Wang and S. Torquato, Designer pair statistics of disordered many-particle systems with novel properties, J. Chem. Phys. **160**, 044911 (2024).